\documentclass[fleqn,usenatbib]{mnras}

\usepackage[T1]{fontenc}

\usepackage{graphicx}	
\usepackage{amsmath}	
\usepackage{amssymb}	
\usepackage{mathrsfs}
\usepackage[english]{babel}
\usepackage{subfig}

\usepackage{hyperref}
\graphicspath{{./images/}}

\usepackage[dvipsnames]{xcolor}

\newcommand{\review}[1]{#1}

\title[VSI parameter study]{High Resolution Parameter Study of the Vertical Shear Instability}

\author[N.Manger et al.]{
Natascha Manger,$^{1,2}$\thanks{E-mail: nmanger@flatironinstitute.org}
Hubert Klahr$^{1}$,
Wilhelm Kley$^{3}$,
and Mario Flock$^{1}$
\\
$^{1}$Max Planck Institute for Astronomy, K{\"o}nigstuhl 17, 69117 Heidelberg, Germany\\
$^{2}$Center for Computational Astrophysics, Flatiron Institute, 162 Fifth Ave, New York, NY 10010, USA\\
$^{3}$Institut f{\"u}r Astronomie und Astrophysik, Universit{\"a}t T{\"u}bingen, Auf der Morgenstelle 10, 72076 T{\"u}bingen, Germany
}

\date{Accepted XXX. Received YYY; in original form ZZZ}

\pubyear{2019}

\begin{document}
\label{firstpage}
\pagerange{\pageref{firstpage}--\pageref{lastpage}}
\citestyle{egu}
\bibliographystyle{mnras}

\maketitle

\begin{abstract}
Theoretical models of protoplanetary disks have shown the Vertical Shear Instability (VSI) to be a prime candidate to explain turbulence in the dead zone of the disk. However, simulations of the VSI have yet to show consistent levels of key disk turbulence parameters like the stress-to-pressure ratio $\alpha$. We aim to reconcile these different values by performing a parameter study on the VSI with focus on the disk density gradient $p$ and aspect ratio $h := H/R$. We use full 2$\pi$ 3D simulations of the disk for chosen set of both parameters. All simulations are evolved for 1000 reference orbits, at a resolution of 18 cells per h. We find that the saturated stress-to-pressure ratio in our simulations is dependent on the disk aspect ratio with a \review{strong} scaling of $\alpha\propto h^{2.6}$\review{, in contrast to the traditional $\alpha$ model, where viscosity scales as $\nu \propto \alpha h
^2$ with a constant $\alpha$.} 
We also observe consistent formation of large scale vortices across all investigated parameters. The vortices show uniformly aspect ratios of $\chi \approx 10$ and radial widths of approximately 1.5 $H$. With our findings we can reconcile the different values reported for the stress-to-pressure ratio from both isothermal and full radiation hydrodynamics models, and show long-term evolution effects of the VSI that could aide in the formation of planetesimals.
\end{abstract}


\defcitealias{Manger+Klahr2018}{MK18}  

\section{Introduction}

The mechanisms governing angular momentum transport in protoplanetary disks are still not well understood. The gas molecular viscosity is orders of magnitude to weak to facilitate the angular momentum transport required to explain the timescales found in observations. To allow nevertheless for a simple description, \citet{Shakura+Sunyaev1973} introduced a parametric turbulent viscosity prescription where the strength of the viscosity depends only on a dimensionless parameter $\alpha$ 
relating the viscous stresses linearly to thermal and magnetic pressure, see eq.~(\ref{eq:alpha}) below.

Many physical explanations yielding sufficiently large values for the $\alpha$-parameter have since been proposed. The first strong candidate was the Magneto-Rotational Instability \citep[MRI,][]{Balbus+Hawley1991}, an instability acting in rotating disks where the equations of ideal Magneto-Hydrodynamics (MHD) are applicable. Since then, further research has shown that the ideal MHD equations are only satisfied in close proximity to the protostar ($< 1$ AU) or high up in the disk atmosphere where gas densities are low and ionisation is created by stellar irradiation. In the gas rich midplane of the disk, non-ideal MHD processes suppress the MRI \citep[e.g.][]{Turner+2014, Dzyurkevich+2010}. Additionally, \citet{Lenz+2019} showed that the turbulent $\alpha \sim 10^{-2}$ generated by the MRI is too high to explain the solid mass distribution in the solar system. However, recent models suggest that magneto-centrifugal winds launched from the disk could facilitate angular momentum transport on the level of $\alpha \sim 10^{-3}$ \citep{2017A&A...600A..75B}.

In recent years, pure hydro-dynamical models have emerged to explain turbulent angular momentum transport in the absence of strong coupling to magnetic fields. Among these are the Convective Overstability \citep[COS,][]{Klahr+Hubbard2014,Lyra2014} and its non-linear cousin Subcritical Baroclinic Instability \citep[SBI,][]{ Klahr+Bodenheimer2003,Petersen+2007a,Petersen+2007b,Lesur+Papaloizou2010}, the Zombie Vortex Instability \citep[ZVI,][]{Marcus+2015,Marcus+2016,Barranco+2018} and the Vertical Shear Instability \citep[VSI][]{Goldreich+Schubert1967,Fricke1968,Nelson+2013}. All these instabilities rely on the fact that baroclinic disks that cool on a certain timescale can become unstable despite them being stable according to the Solberg-Hoiland criteria \citep{Ruediger+2002,Urpin2003,Arlt+Urpin2004}. Each of these instabilities works in a specific ranges of cooling times: The ZVI requires the disk to cool on very long time-scales ($\tau_\mathrm{cool} \gg 1/\Omega$), whereas the COS and SBI need the cooling time to be on the order of $\tau_\mathrm{cool} \sim 1/\Omega$, and the VSI requires the cooling of the disk to be very fast ($\tau_\mathrm{cool} \ll 1/\Omega$), or even instantaneous \citep{Lin+Youdin2015}. Both the VSI and the COS have been shown to facilitate angular momentum transport at the level of $\alpha \sim 10^{-5} - 10^{-3}$ \citep{Raettig+2013,Lyra2014,Stoll+Kley2014,Stoll+Kley2016,Flock+2017,Manger+Klahr2018}.

\citet{Richard+2016} showed the VSI to be capable to spawn small, short lived vortices in localized disk models, identified by \citet{Latter+Papaloizou2018} as generated by a parasitic Kelvin-Kelmholtz instability. In \citet{Manger+Klahr2018}, we showed the VSI to be also capable to generate large scale, long lived vortices via the Rossby-Wave-Instability \citep[RWI,][]{Lovelace+1999,Li+2000,Li+2001}. These vortices are high pressure maxima and are therefore able to trap dust particles efficiently \citep{Weidenschilling1977,Barge+Sommeria1995}, making them prime locations for planetesimal formation via gravo-turbulent mechanisms. \citet{Ricci+2019} showed that these vortices should be detectable with ALMA and ngVLA.

In this work, we present the results of the first high resolution 3D parameter study conducted for protoplanetary disks with active Vertical Shear Instability. Because full 3D simulations are resource consuming, we focus specifically on the parameters $p$ and $h$ controlling the radial density gradient and the disk aspect ratio, respectively. Because we are investigating the behaviour of the VSI within a controlled parameter set, we choose to employ a simplified cooling law in favour of a more physically correct radiative model \citep[e.g. as in ][]{Stoll+Kley2016,Flock+2017}, because it allows us the freedom to set uniform cooling parameters throughout the disk. To make the simulations comparable, we choose the numerical setup to achieve a resolution of approximately 18 grid cells per scale height in all directions in all simulations. 

Our goals in this study are three-fold: 
\begin{enumerate}
    \item Investigation on whether the results obtained in previous 2D axisymmetric studies of the VSI \citep{Nelson+2013,Richard+2016} are applicable to full 3D disks. We will show that the $\alpha$-value generated by the VSI turbulence scales with the disk's aspect ratio as $\alpha\propto h^{2.6}$, while being otherwise consistent with previous works.
    \item Reconciliation of differences in turbulence parameters found when comparing recent work on full 3D simulations utilising different disk parameters \citep[e.g][]{Stoll+Kley2014,Flock+2017,Manger+Klahr2018}.
    \item Further investigation into vortex formation within the framework of the VSI and it's relation to the disk parameters. We show that vortex formation is ubiquitous in VSI turbulent disks and that the vortex size is related to the disk aspect ratio.
\end{enumerate}

This paper is structured as follows: In section 2 we briefly describe the methods and initial conditions used in this study. In section 3 we present our simulation results, and in section 4 we take a closer look at the vortices formed in the disk. Section 5 presents theoretical explanations to our findings and connections to previous studies. Finally, in section 6, we summarise our results and present conclusions to our work.

\section{Model}
\label{sec:model}
As in our previous work \citet[][hereafter MK18]{Manger+Klahr2018} we study a three-dimensional section of a protoplanetary disk numerically solving the equations of ideal hydrodynamics with a specific cooling prescription as given below. 
We use a setup similar to the one used in \citetalias{Manger+Klahr2018}. The initial conditions are defined in force equilibrium, where the density is given by
\begin{equation}
\rho = \rho_0 \left(\frac{R}{R_0}\right)^p \exp\left(-\frac{Z^2}{2\,H^2}\right) \, ,
\end{equation}
and the initial angular velocity of the disk by
\begin{equation}
v_\phi = \Omega_\mathrm{K}\,R \left[1+q-\frac{q\,R}{\sqrt{R^2+Z^2}}+(p+q)\left(\frac{H}{R}\right)^2\right]^{\frac{1}{2}} \qquad . \label{eqn:Omega}
\end{equation}
While in the simulations we use spherical polar coordinates, $R$ and $Z$ in the above equations denote cylindrical coordinates, and $H(r)$ is the half thickness (vertical pressure scale height) of the disk.
We use an ideal equation of state $\rho e = \frac{P}{\gamma -1}$ with the pressure defined as $P = c_\mathrm{s}^2 \rho$. Here, the isothermal sound speed, $c_\mathrm{s}^2 = c_0^2 \left(\frac{R}{R_0}\right)^q $, is given as a function of radius, where we choose $q=-1$ in all our simulations. All other quantities used are defined in table \ref{tab:symbols}. 

\begin{table}
\renewcommand\arraystretch{1.5}
\begin{tabular}{l l l}
\hline
Symbol & Definition & Description \\
\hline
$ R,\phi,Z $&  & cylindrical coordinates \\
$ r,\theta,\varphi $& & spherical coordinates\\
$ R_0,Z_0 $& & cylindrical reference coordinates \\
$\rho $& & density \\
$\rho_0$& & reference density \\
$ H$& $ =\frac{c_s}{\Omega_K} $ & disk pressure scale height \\
$h$ &$ = \frac{H}{R} = \frac{c_\mathrm{s}}{v_{\rm K}}$ & disk geometric scale height\\
$P$ &$ = c_\mathrm{s}^2 \rho $ &  pressure\\
$c_\mathrm{s}$ & $ = c_0 \left(\frac{R}{R_0}\right)^{q/2}  $ & isothermal sound speed \\
$c_0$ & $ = \frac{k T_0}{\mu m_H} $ & reference sound speed  \\
$p $& $=\frac{\mathrm{d}\log \rho}{\mathrm{d}\log R}$ & radial density slope \\
$q $& $=\frac{\mathrm{d}\log T}{\mathrm{d}\log R}=-1$ & radial temperature slope \\
$e $ & & specific internal energy density \\
$\gamma $ &$= 1.44 $ & adiabatic index \\
$v_\phi $& & azimuthal velocity \\
$v_\mathrm{K}$& $ = \sqrt{\frac{G M_\star}{R}} $ & Kepler azimuthal velocity \\
$\Omega_\mathrm{K}$ & $ = v_K R^{-1} $ & Kepler angular frequency \\
$\tau_\mathrm{relax}$& & temperature relaxation time \\
$ T_{r,\phi}$ & eqn. \ref{eqn:ViscStressTens} & r$\phi$ component of viscous stress tensor\\  
$\alpha$ & $ = \frac{ T_{r,\phi}}{P}$ & turbulence parameter\\
$v_\mathrm{rms}$ & eqn. \ref{eqn:vrms} & root mean squared velocity\\
$\mathcal{E}$ & eqn. \ref{eqn:SpKinEnFreq} & spectral specific kinetic energy \\
$\tilde{v}_i $ & eqn. \ref{eqn:FourierTransf} & velocity in i direction in fourier space\\
$\omega$ & $= \nabla \times \vec{v}$ & vorticity\\
$\mathscr{L}$ & eqn. \ref{eqn:RWI} & RWI criterion\\
\hline
\end{tabular}
\caption{List of all symbols used within this work.}
\label{tab:symbols}
\end{table}

Because we are interested in the dynamics of the VSI with changing cooling law, we use a simple prescription with
\begin{equation}
\frac{dP}{dt} = -\frac{P-\rho \,c_\mathrm{s,init}^2}{\tau_\mathrm{relax}}
\end{equation}
where, to force the disk to be approximately isothermal, we define the relaxation time $\tau_\mathrm{relax}$ equal to the simulation time step $dt\sim 10^{-4}\frac{2 \pi}{\Omega_\mathrm{K}}$. The influence of longer cooling times on the VSI will be explored in a future publication.

For our computations we use the multi-purpose MHD Godunov code \textsc{PLUTO} \citep{Mignone+2007} with the hllc solver \citep{Toro2009}. We use the piecewise parabolic method (PPM) by \citet{Mignone2014} for the spatial reconstruction and a 3rd-order Runge-Kutta time integration.
All simulations performed are listed in table \ref{tab:simParam} with their parameters. We choose the grid sizes listed in column 3 of table \ref{tab:simParam} to ensure all simulations run share a common resolution of circa 18 cells per $H$ in all directions.

We use periodic boundary conditions in azimuthal direction and modified outflow conditions in vertical direction, where we ensure zero inflow into the domain and additionally extrapolate the Gaussian density profile into the ghost zones. In radial direction, we use reflective boundaries combined with buffer zones, where we relax all variables to their initial values. These buffer zones are excluded in our analysis of the simulations.

\begin{table*}
\renewcommand\arraystretch{1.5}
\centering
\begin{tabular}{c c c l l c c }
\hline
model & $r_\mathrm{in, out}/R_0$ & grid size ($\mathrm{N}_\mathrm{r}\times\mathrm{N}_\theta \times \mathrm{N}_\varphi$) & $p$ & $h$ & $\langle \alpha \rangle /10^{-4}$ & $\Gamma \left[ 2\pi/\Omega \right]$ \\ \hline
p0.6h0.1  & $0.5-2.0$ & 256$\times$128$\times$1024 & $-0.66$ & $0.1$ & $8.4\pm 2.6$ & $0.36$ \\
p1.5h0.1  & $0.5-2.0$ & 256$\times$128$\times$1024 & $-1.5$ & $0.1$ & $9.5\pm 2.1$ & $0.38$\\
p1.5h0.07 & $0.6-1.6$ & 256$\times$128$\times$1464 & $-1.5$ & $0.07$ & $2.7\pm 0.6$ & $0.3$\\
p0.6h0.05 & $0.7-1.4$ & 256$\times$128$\times$2048 & $-0.66$ & $0.05$ & $1.5\pm 0.3$ & $0.2$\\
p1.5h0.05 & $0.7-1.4$ & 256$\times$128$\times$2048 & $-1.5$ & $0.05$ & $1.2\pm 0.2$ & $ 0.2$\\
p1.5h0.03 & $0.8-1.2$ & 256$\times$128$\times$3402 & $-1.5$ & $0.03$ & $0.5\pm 0.2 $ & $0.08$\\
\hline
\end{tabular}
\caption{List of simulation and model parameters. From left to right: model name, radial domain size, numerical grid size, density slope parameter, disk aspect ratio and space and time averaged stress to pressure value. The vertical and azimuthal domain sizes for all simulations are $\theta = \pm 3.5 H$ and $\phi = 0 - 2\pi$, respectively.
}
\label{tab:simParam}
\end{table*}

\section{Analysis of the Disk Gas Kinematics}
\label{sec:results1}
In this section, we present the results of our parameter study on the influence of radial density gradient and disk aspect ratio on the vertical shear instability, focusing particularly on the angular momentum generating stresses and gas rms-velocities.

\subsection{Stress-to-Pressure ratio}
To measure the strength of the VSI turbulence in the disk, we calculate the $T_{r,\phi}$ component of the Reynolds tensor
\begin{equation}
T_{r,\phi} \equiv \langle \rho \delta v_r \delta v_\phi \rangle =
\langle \rho v_r v_\phi \rangle - \langle \rho v_r \rangle \langle v_\phi \rangle \,, \label{eqn:ViscStressTens}
\end{equation}
where $\langle ~ \rangle$ denotes an average in azimuth.
$T_{r,\phi}$ measures the strength of angular momentum transport generated by the disk turbulence. To present the values in a non-dimenional fashion, we normalise $T_{r,\phi}$
by the azimuthally averaged pressure to obtain $\alpha$ as defined by \citet{Shakura+Sunyaev1973}:
\begin{equation}
\label{eq:alpha}
\alpha = \frac{ T_{r,\phi}}{\langle P \rangle} \qquad.
\end{equation}

\begin{figure}
\centering
\subfloat{\includegraphics[width=\columnwidth]{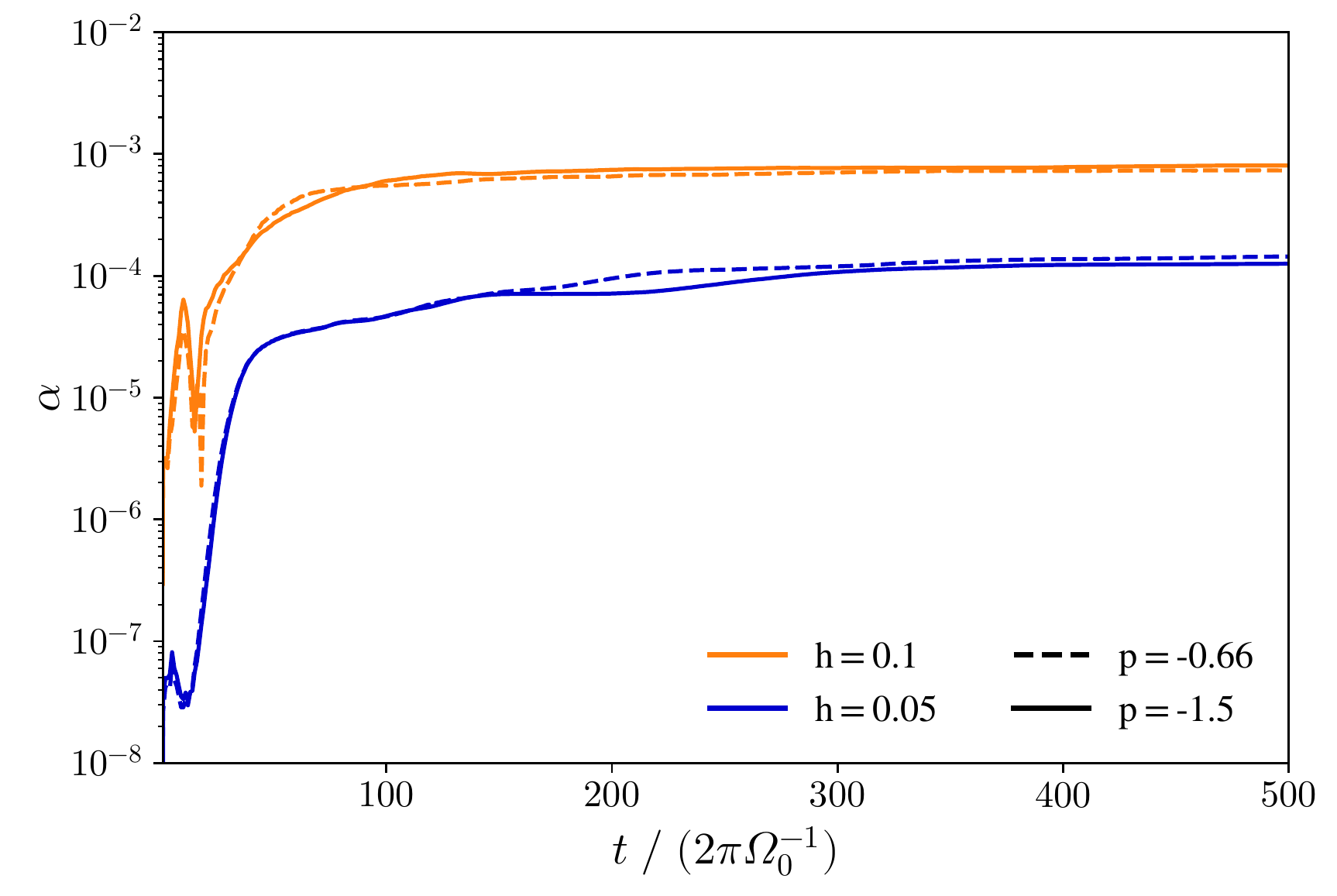}}\newline
\subfloat{\includegraphics[width=\columnwidth]{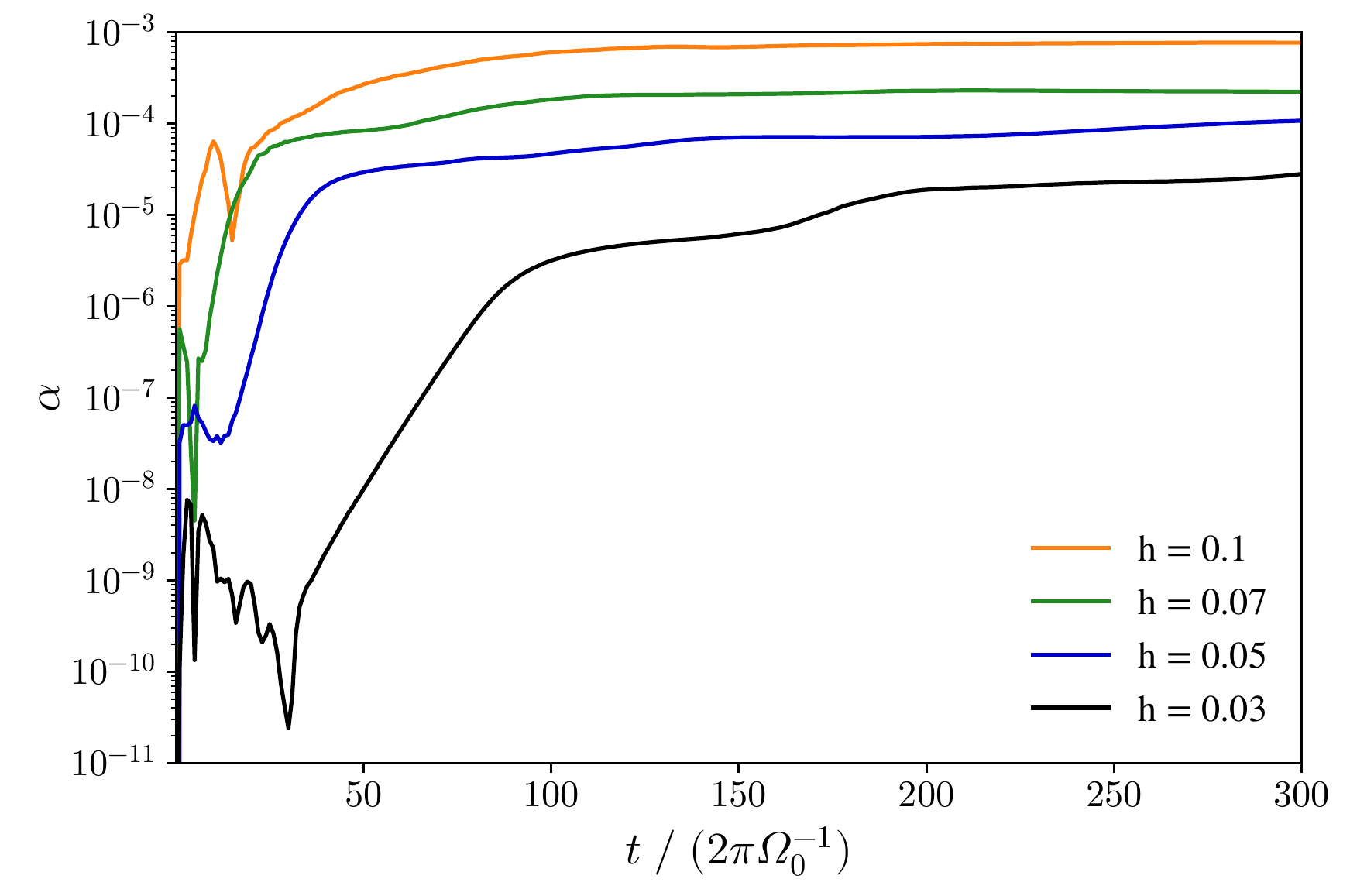}}
\caption{Cumulative space and time average of the stress-to-pressure ration for simulations with different density slopes $p$ (top) and aspect ratios $h$ at constant $p=-1.5$ (bottom).}
\label{fig:alphaTs}
\end{figure}

\begin{figure}
\centering
\subfloat{\includegraphics[width=\columnwidth]{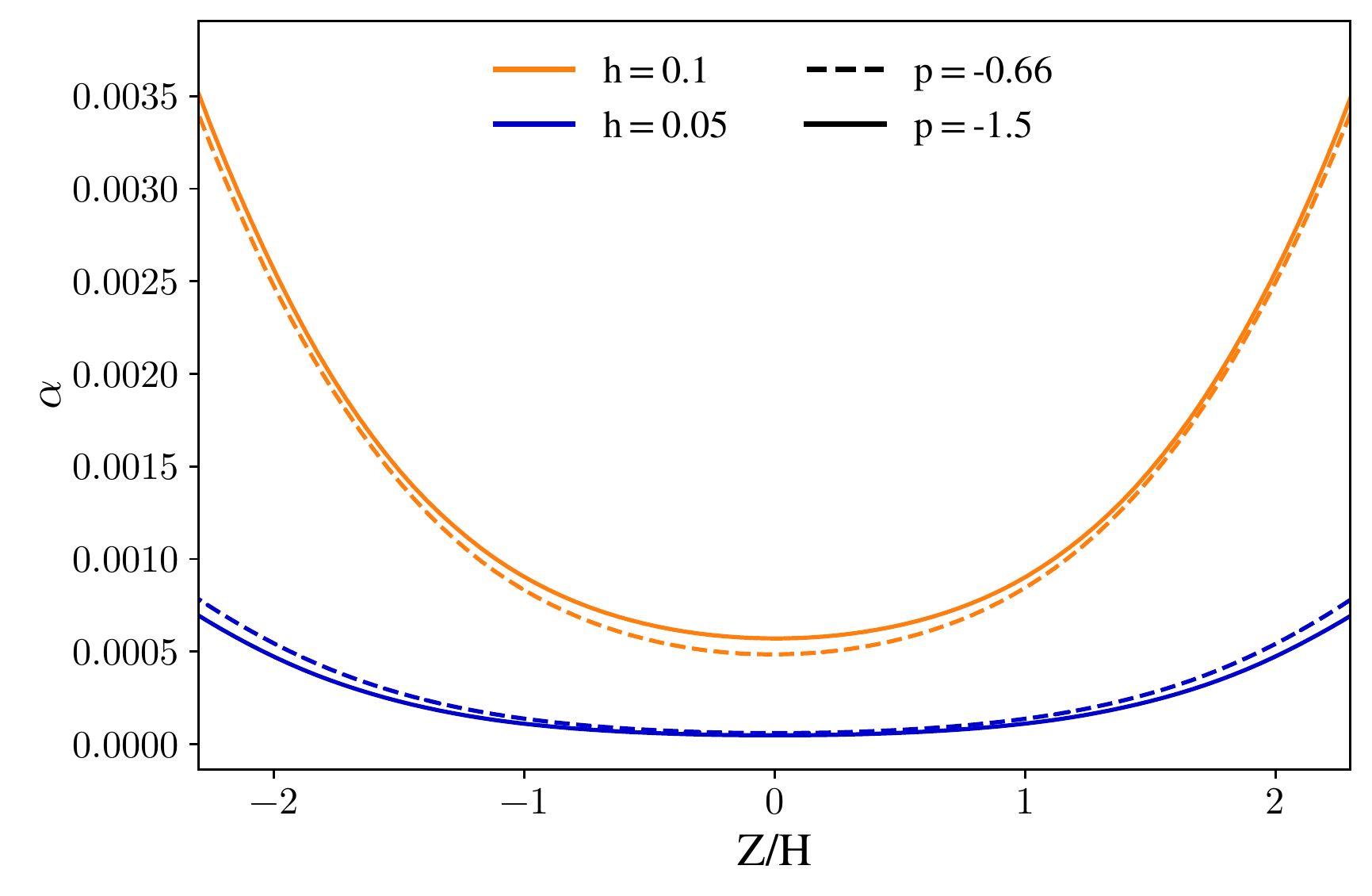}}\\
\subfloat{\includegraphics[width=\columnwidth]{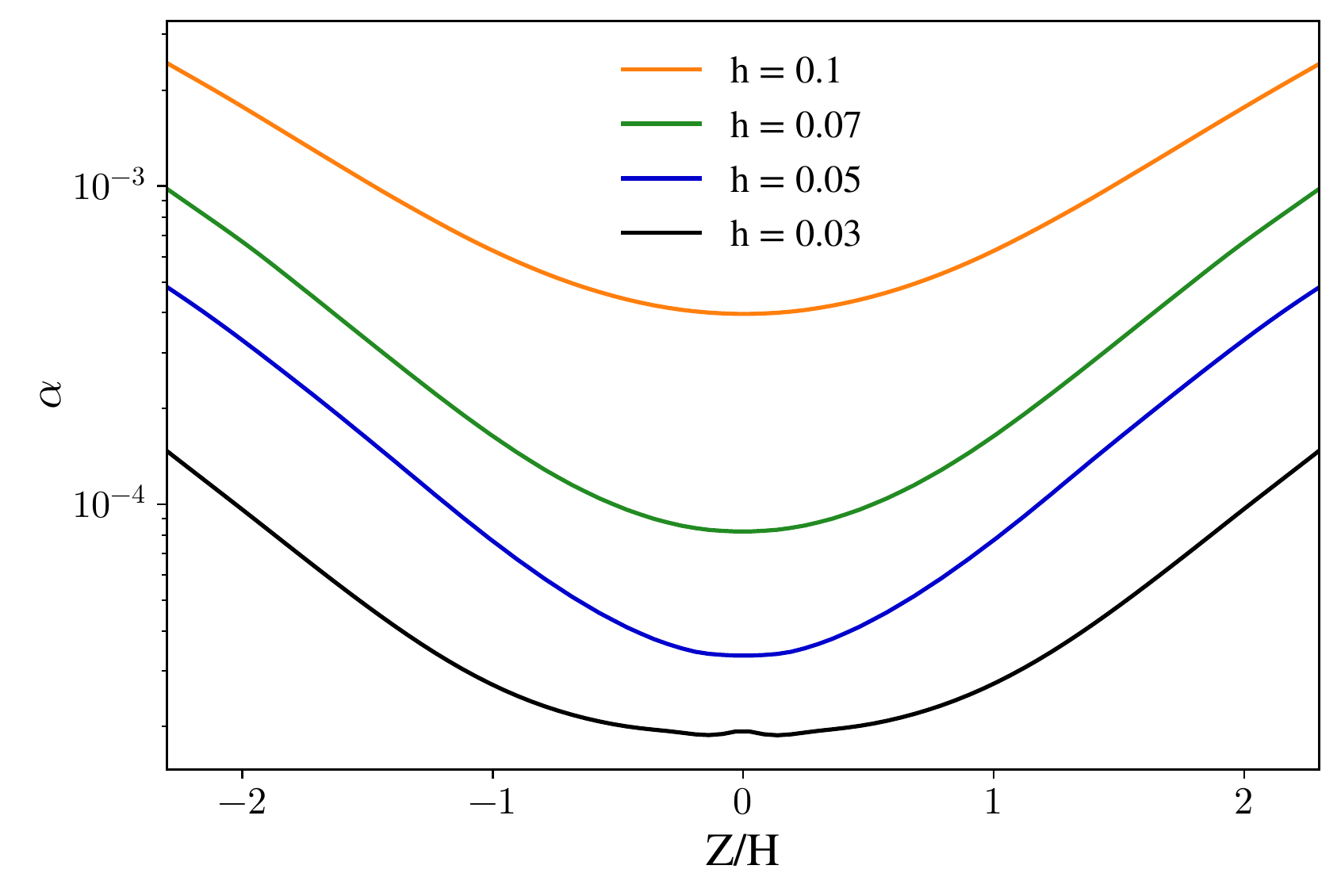}}
\caption{Stress-to-pressure ratio as a function of height. The top panel shows the dependence on the density slope $p$ for two values of $h$. The bottom panel shows the dependence on disk aspect ratio $h$ for a slope $p=-1.5$. The values are averaged in radial and azimuthal direction and from 600-1000 orbits.
}
\label{fig:alphaZ}
\end{figure}
In figure \ref{fig:alphaTs} we present the cumulative space and time averaged values of $\alpha$. The top panel of figure \ref{fig:alphaTs} compares simulations with two different density slopes $p=-0.66$ and $p=-1.5$ at two different values of the disk aspect ratio $h=0.05$ and $h=0.1$, whereas the bottom panel compares simulations with four different values of $h$ and a constant value of $p=-1.5$. 

The time evolution of the simulations with $h=0.1$ shows a rapid growth of the $\alpha$ value in the first few tens of orbits of the simulation, after which a slower growth phase leads to growth to the final saturated phase of the turbulence after around 100 orbits, where values of $\alpha = 9 \cdot 10^{-4} $ are reached. These values are comparable the ones we reported for our simulations in \citetalias{Manger+Klahr2018} for simulations with lower azimuthal but similar radial and meridional resolution. A similar behaviour is observed for the simulations with $h=0.05$, which also show a first strong growth phase up to around 50 orbits, after which slower growth phase is observed until around 300 orbits, where a steady state value of $\alpha = 1\cdot 10^{-4}$ is reached.

Comparing all four simulation runs, we find that the density slope $p$ does not significantly influence the average value of the turbulent stresses. The comparison however shows a clear correlation of the disk aspect ratio with the turbulent $\alpha$ values of the disk. This result has been expected, as the disk aspect ratio is proportional to the disk temperature and therefore a larger value of $h$ leads to a larger overall temperature in the disk and to stronger turbulent velocities as visible in figure \ref{fig:vrmsTs}. 
We therefore ran additional simulations with aspect ratios $h=0.07$ and $h=0.03$, presented in the lower panel of figure \ref{fig:alphaTs} along the simulations discussed above. In comparing these four simulation, we see a clear trend with disk aspect ratio emerging: Simulations with larger $h$ show overall stronger turbulent angular momentum transport, with $\alpha = 9\cdot 10^{-4}$ for $h=0.1$ going down to $5 \cdot 10^{-5}$ for $h=0.03$. A full list of averaged saturated $\alpha$-values is listed in \ref{tab:simParam}, where the errors listed are calculated for fluctuations in time only.
The additional simulations also confirm the trend for later onset of turbulent growth and longer times until saturation. 

Figure \ref{fig:alphaZ} shows the dependence of $\alpha$ on height above the midplane, where the top panel again compares two different density slopes at $h=0.05$ and $h=0.1$, and the bottom panel for different $h$ at $p=-1.5$. The values are averaged in radial and \review{azimuthal} direction and between 600 and 1000 reference orbits.
We again find no evidence that the initial density gradient p influences the generated stresses. For both $h=0.05$ and $h=0.1$ the deviations between the curves representing $p=-0.66$ and $p=-1.5$ are minor and can be explained with statistical effects. In both the top and bottom panel the trend of overall increasing $\alpha$ with increasing $h$ is observed. We also find that higher $h$ leads to a larger difference between the $\alpha$ values in the midplane and the upper disk layers, though the general dependence of $\alpha$ with height described in \citetalias{Manger+Klahr2018} is found in all simulations.  This is also in accordance with simulations presented by e.g. \citet{Stoll+Kley2014}, \citet{Stoll+Kley2017} and \citet{Flock+2017}.

In Appendix A, we present a resolution study of the azimuthal direction. We find that the choice of resolution in this direction is not influencing the time evolution of the $\alpha$-value, even at one-eighth the fiducial resolution. This is likely due to the axisymmetric nature of the initial linear VSI growth. However, at the lowest resolution the vertical profile of $\alpha$ becomes steeper, as the resolution is only 2 cells per $H$.

\subsection{rms-velocities}
We now take a look at the initial growth phase of the VSI by calculating the rms-velocity defined as
\begin{equation}
v_\mathrm{rms}= \sqrt{\left(v_r -\langle v_r\rangle\right)^2 + \left(v_\theta -\langle v_\theta\rangle\right)^2+ \left(v_\varphi -\langle v_\varphi\rangle\right)^2} \: , \label{eqn:vrms}
\end{equation}
with brackets representing spatial averages in $\phi$ direction. The averages $\langle v_r \rangle$ and $\langle v_\theta\rangle$ are assumed equal to zero.

\begin{figure}
    \centering
\subfloat{\includegraphics[width=\columnwidth]{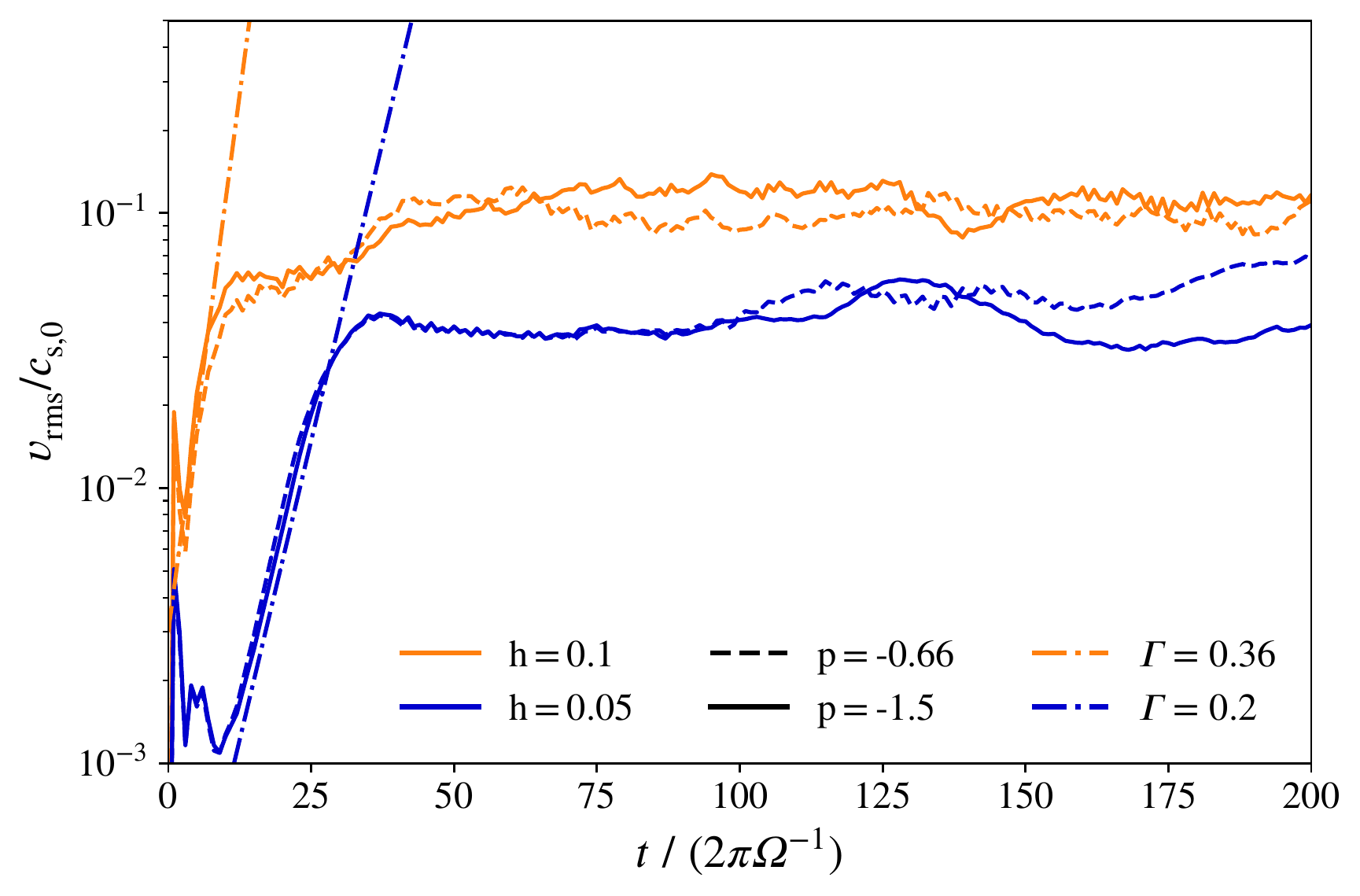}}\\
\subfloat{\includegraphics[width=\columnwidth]{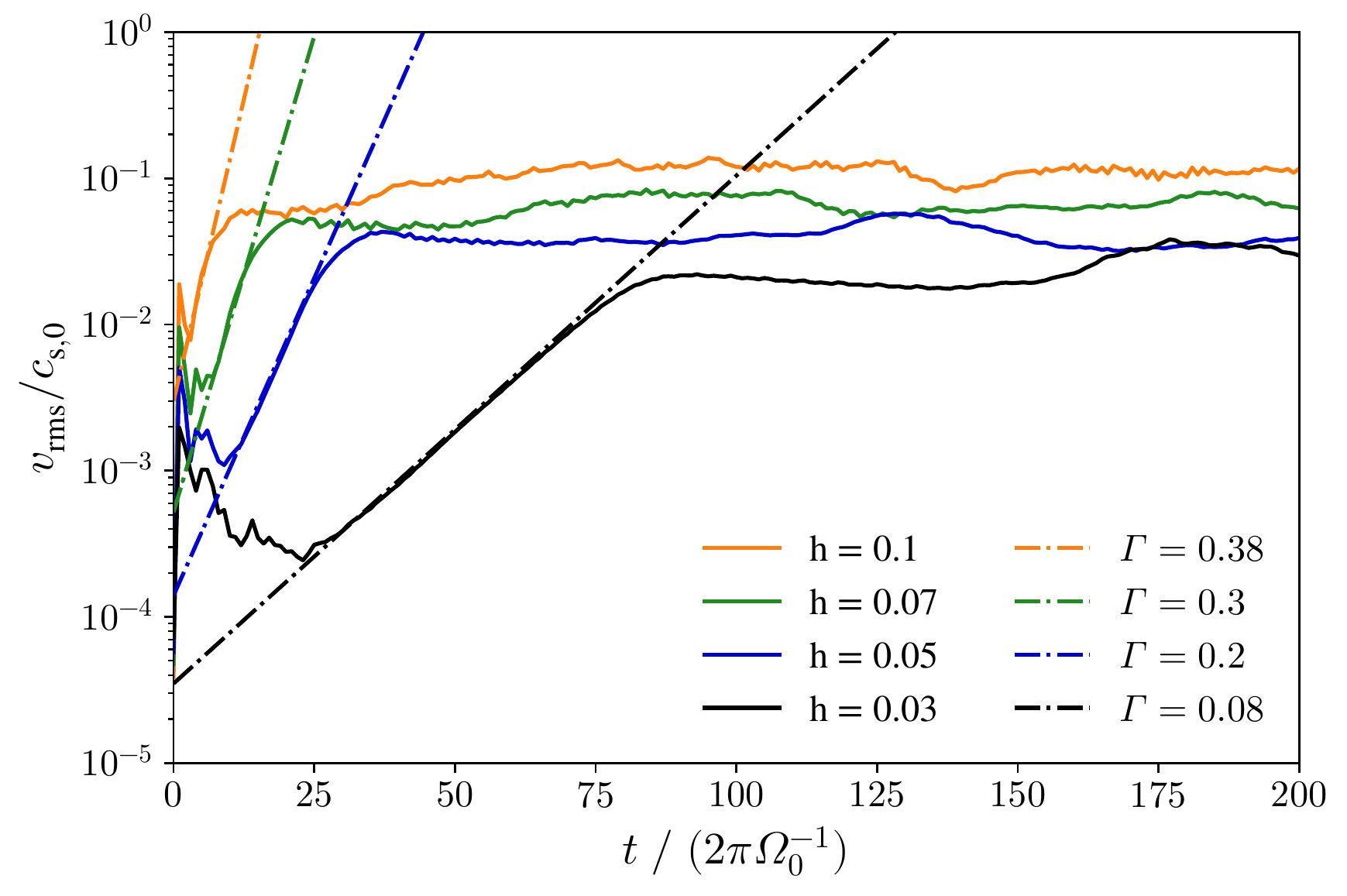}}
\caption{Rms-velocities normalised by reference sound speed measured for the initial growth phase of the instability. The top panel compares the disks with $p=-0.66$ and $p=-1.5$ at $h=0.05$ and $h=0.1$, while the bottom panel compares different $h$ from 0.03 to 0.1 at a common $p$ value. We additionally plot curves with exponential growth rates $\Gamma$ for comparison.}
\label{fig:vrmsTs}
\end{figure}

\begin{figure}
    \centering
\subfloat{\includegraphics[width=\columnwidth]{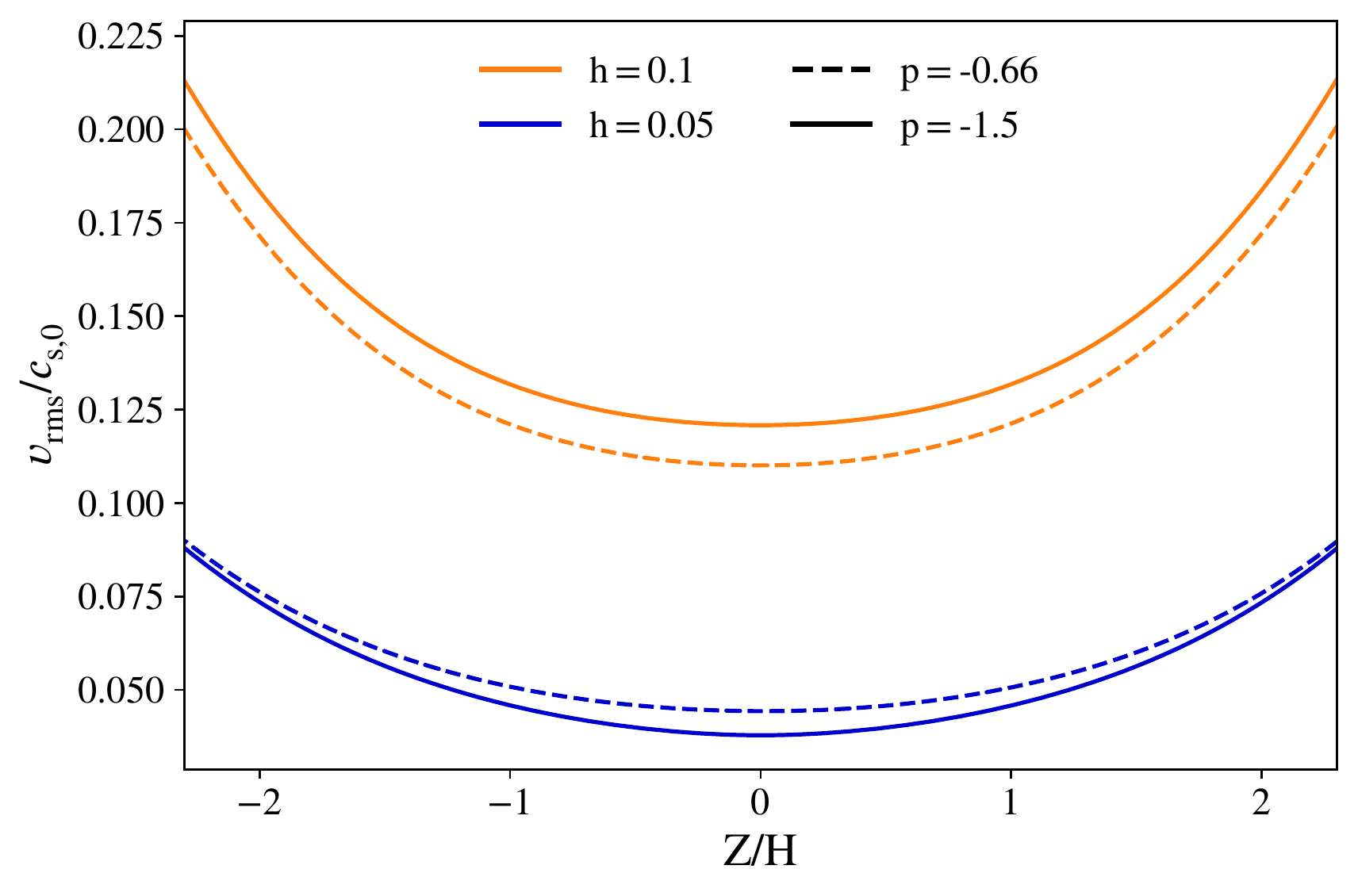}}\\
\subfloat{\includegraphics[width=\columnwidth]{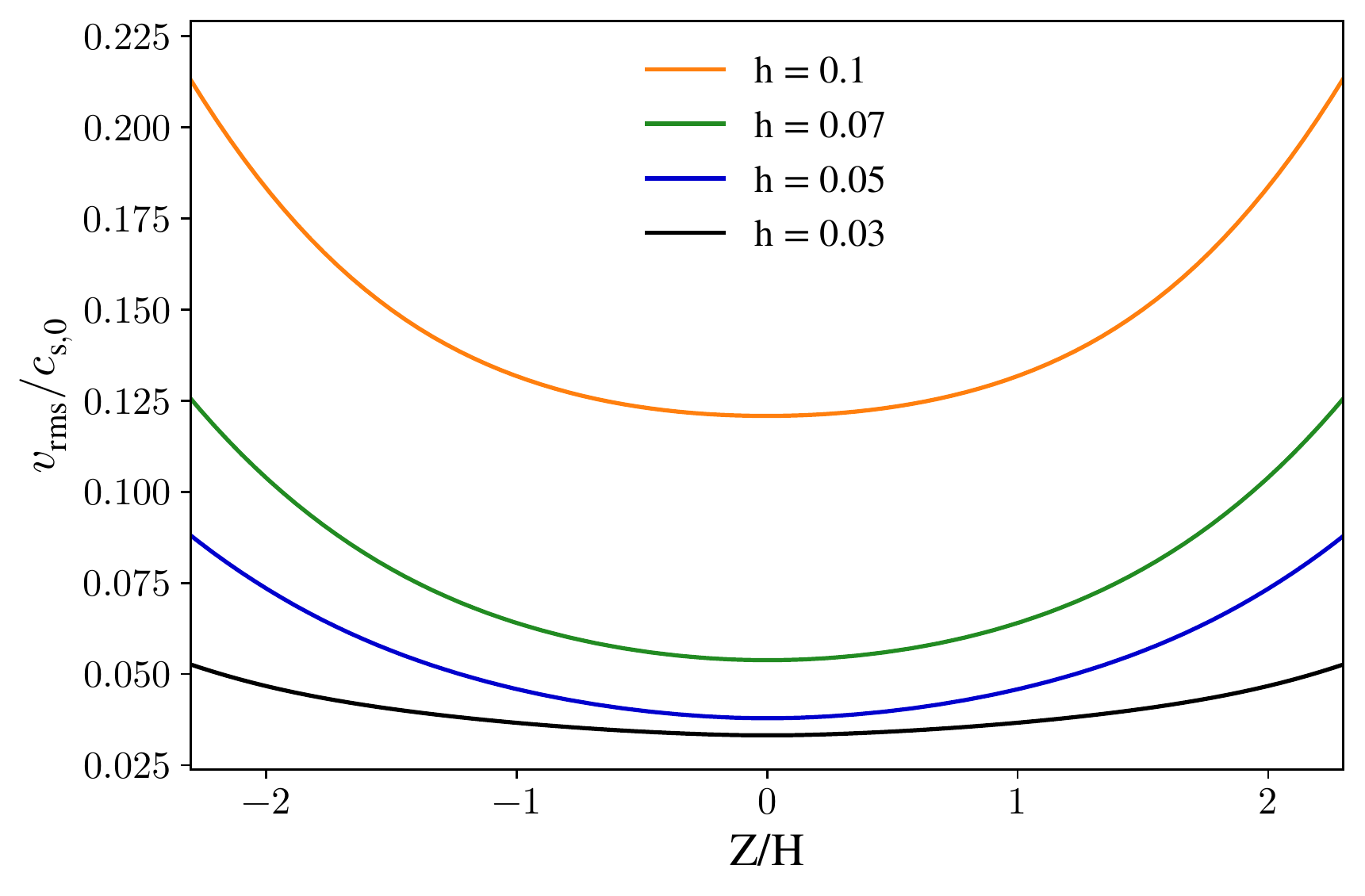}}
    \caption{Rms-velocities as a function of distance from the disk midplane. The averages are calculated for the remaining dimensions and over the simulation time. Simulations shown in each panel ar the same as in figure \ref{fig:vrmsTs}.}
    \label{fig:vrmsZ}
\end{figure}

In figure \ref{fig:vrmsTs} we plot the spatial average of the rms-velocity as a function of time for the first 200 orbits of the simulations. We find that the simulations for $h=0.1$ with different values of p start their growth at a similar time and with identical growth rates of $\Gamma=0.36$ per orbit. We also observe the secondary growth phase seen already in the time evolution of the $\alpha$ value.  A similar qualitative behaviour is found for the simulation with $h=0.05$. The simulations both settle to a steady state value of $v_\mathrm{rms} \approx 0.1 c_\mathrm{s}$. The onset of the VSI growth is later than for the $h=0.1$ cases in agreement with the results from the analysis of the $\alpha$ stresses. The second growth phase is however not observed in the simulations with $h=0.05$, instead we find a plateau in the $v_\mathrm{rms}$ values after the initial growth phase and further growth is only observed after additional 50 orbits. For both values of $p$ we find the initial growth rate to be $\Gamma=0.2$ per orbit.

In the bottom panel of figure \ref{fig:vrmsTs} we present the rms-velocities for the simulations with $p=-1.5$ and different values of $h$. For the simulation with $h=0.07$ we find a growth rate of $\Gamma=0.3$ which is in good agreement with the linear scaling predicted by analytical calculations \citep{Nelson+2013}. For $h=0.03$ we however get $\Gamma = 0.08$, which is lower than  we would expect based on the growth rates of the other simulations in combination with a linear scaling, from which a growth rate of $\Gamma=0.12$ would be expected. A possible explanation is that the simulation with $h=0.03$ was run using the FARGO scheme to increase computational efficiency, which resulted in a time step time step twice than in the other simulation. As the cooling time in our setup is linked to the simulation time step, the simulation thus has a slightly longer cooling time, possibly influencing the VSI growth.

In figure \ref{fig:vrmsZ} we display the vertical dependence of the rms-velocities. In the top panel, we again find the curves corresponding to the same $h$ to agree well with each other, supporting our claim that the initial density gradient does not influence the simulation result. For $h=0.1$ the rms-velocity at the midplane is $v_\mathrm{rms} = 0.13 c_\mathrm{s}$ and rises to $v_\mathrm{rms} = 0.2 c_\mathrm{s}$ in the upper layers. These results are consistent with the results obtained in \citetalias{Manger+Klahr2018} taking into account that we included the $\phi$ component of velocity in the calculation of $v_\mathrm{rms}$ presented in this work.
Comparing the results for different $h$ value shown in the bottom panel of figure \ref{fig:vrmsZ}, we find that the overall shape of the $v_\mathrm{rms}$ stays similar for all cases, although the difference between the velocity at $z=0$ and $z=2H$ increases with increasing $h$. A similar correlation exists between $h$ and the $v_\mathrm{rms}$ value at $z=0$, which is larger for higher $h$. This is expected, as a warmer disk has more shear energy available to be converted into turbulence.

\begin{figure}
    \centering
    \includegraphics[width=\columnwidth]{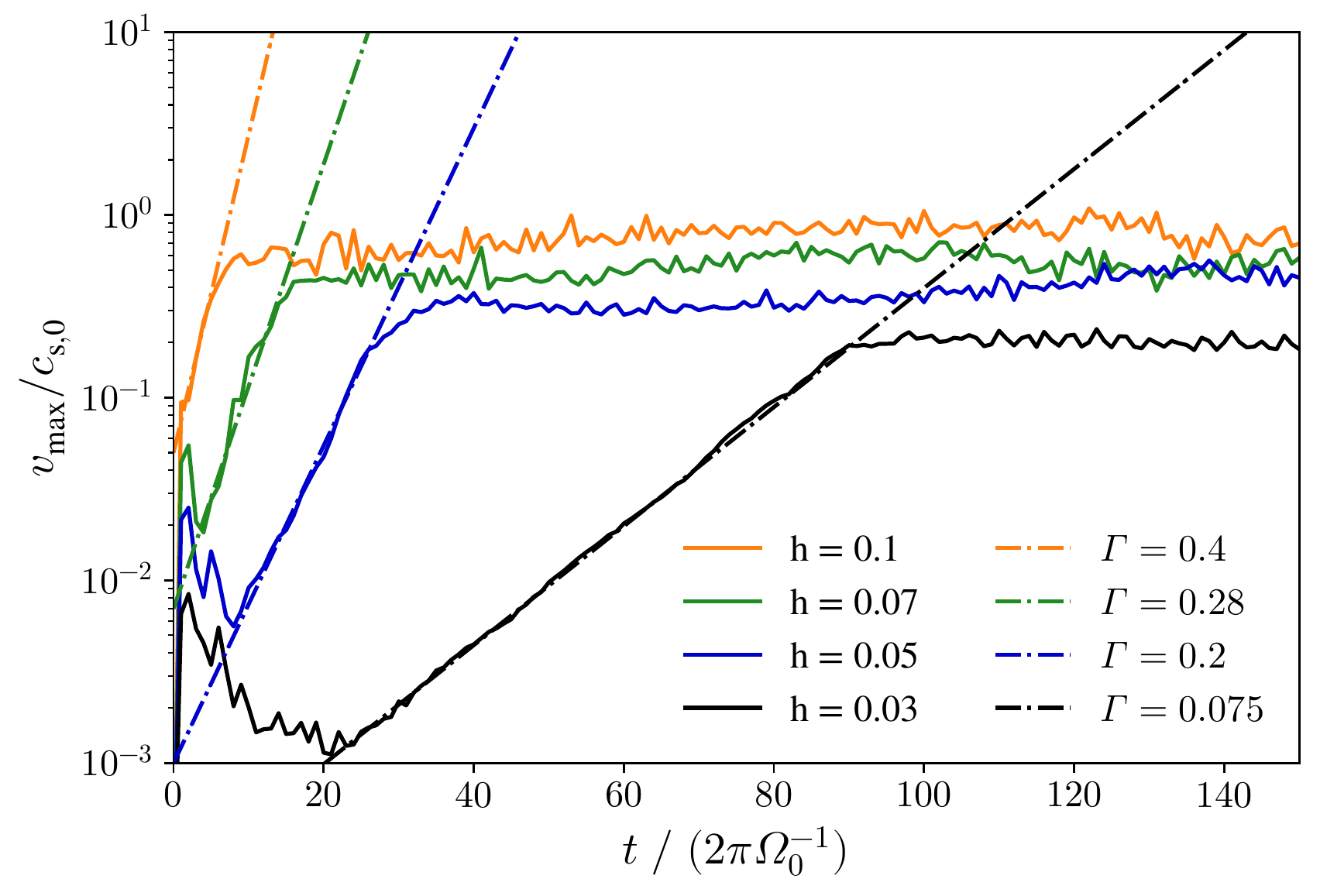}
    \caption{Maximum perturbed velocity as a function of time for the initial growth phase for different values of h. We again show exponential growth rates as a comparison.}
    \label{fig:vmaxTs}
\end{figure}

Because of the vertical variation of the velocity fluctuations and the fact that the growth of an instability is governed by the largest and not the root-mean-square velocity, we calculate the maximum velocity as
\begin{equation}
    v_\mathrm{max} = \sqrt{v_r^2+v^2_\theta} \,.
\end{equation}
Here, we consider only the poloidal motion and neglect the $\phi$ component of velocity in this case as the VSI has been shown to grow axisymmetrically. The results are shown in figure \ref{fig:vmaxTs} for the simulations with $p=-1.5$, where we again plot growth rates for comparison. We find the growth rates obtained using $v_\mathrm{max}$ to be in good agreement with the ones obtained using $v_\mathrm{rms}$.

\subsection{Specific kinetic energy spectrum}

\begin{figure}
    \centering
    \includegraphics[width=\columnwidth]{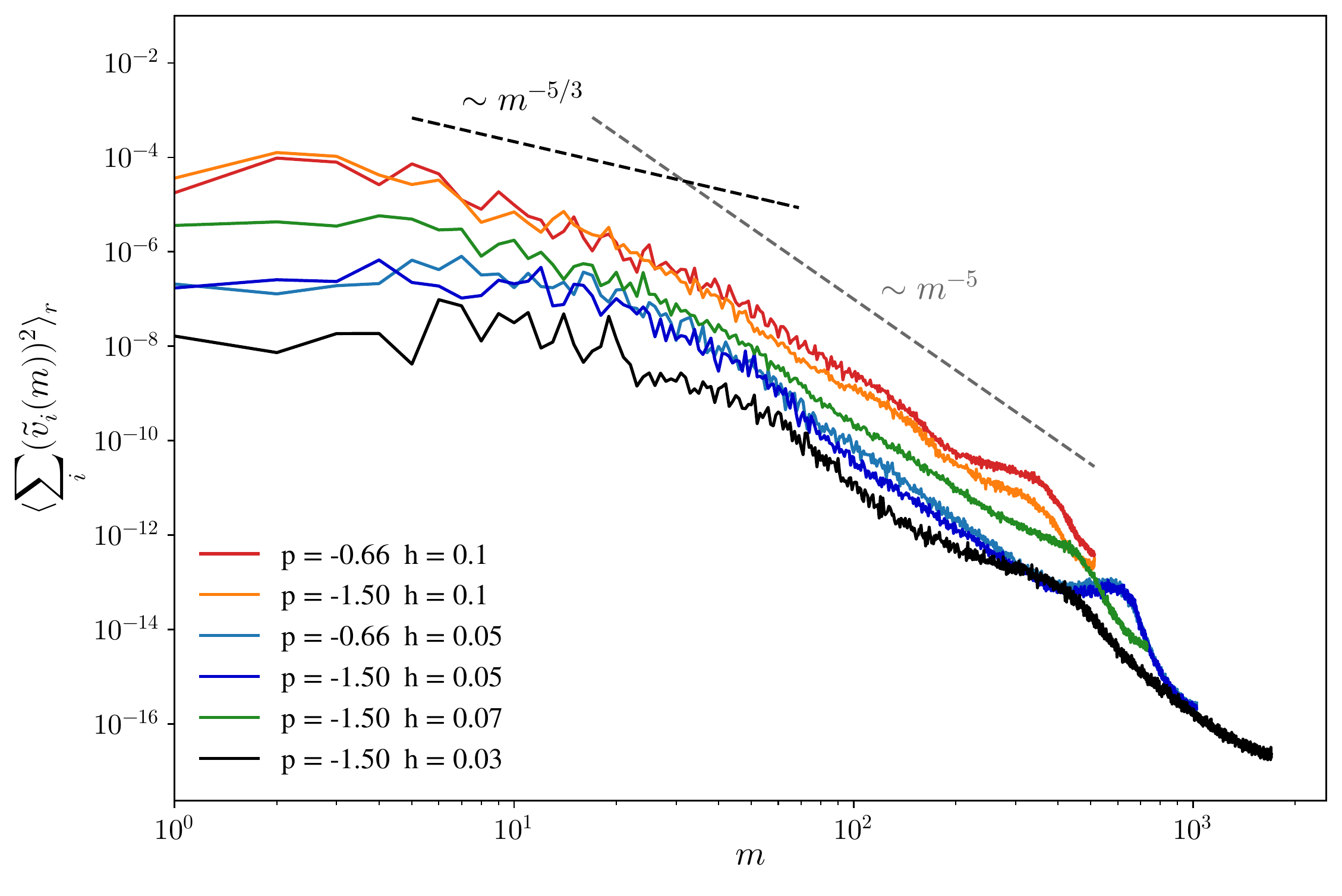}
    \caption{Specific kinetic energy spectrum as a function of azimuthal wavenumber. For comparison, we show both the $m^{-5/3}$ behaviour expected for 3D isotropic Kolmogorov-like turbulence and the closest matching slope $m^{-5}$.}
    \label{fig:ekin_m}
\end{figure}

To gain additional insight into the turbulent behaviour of the VSI we look at the specific kinetic energy spectrum defined as
\begin{equation}
    \mathcal{E}(m) = \langle\lvert\tilde{v}_R\rvert^2 + \lvert\tilde{v}_\theta\rvert^2 +\lvert\tilde{v}_\phi\rvert^2 \rangle_R
    \label{eqn:SpKinEnFreq}
\end{equation}
where $m$ is the azimuthal wave number and $\tilde{v}_k$ is the Fourier transform of the $k$ component of velocity in the azimuthal direction defined by
\begin{equation}
    \tilde{v}_k(m)  = \frac{1}{2\pi} \int\limits_{-\infty}^\infty v_k(\phi)\,e^{-i m \phi}\, \mathrm{d}\phi \qquad.
    \label{eqn:FourierTransf}
\end{equation}
\review{We plot the results as a function of azimuthal wavenumber in figure \ref{fig:ekin_m}. We find that the azimuthal kinetic energy spectrum is not consistent with a 3D Kolmogorov turbulent energy cascade, which would be $\propto m^{-5/3}$. This was expected, as the VSI produces weak turbulence that is always and at all scales dominated by the disk rotation. This means, if the injection scale has a Rossby-number Ro (defined as the ratio of inertial to coriolis forces) of smaller than unity, then the turbulence will always be in the $\mathrm{Ro} <1$ regime at all scales and will not develop isotropic 3D turbulence at any scale \citep{Sharma+2018}. Instead, the results show that the spectrum is proportional to $ m^{-5}$ with a turn-over into a shallower power law at low $m$, which is consistent with an upward cascade transporting energy to larger scales combined with an downward cascade transporting enstrophy, a quantity related to the square of vorticity, to smaller scales \citep{Lyra+Umurhan2019}.} This behaviour is characteristic of 2D turbulence \citep{Kraichnan1971} and is consistent with findings from our previous work presented in \citetalias{Manger+Klahr2018}, although for lower h this behaviour is less clearly visible. We also find the maxima in the spectrum for $m=2-6$ reported previously, although these are also less clearly visible for lower h. We also observe $h=0.03$ differ from this behaviour, here the energy is deposited at $m\geq 6$.

\begin{figure}
    \centering
    \includegraphics[width=\columnwidth]{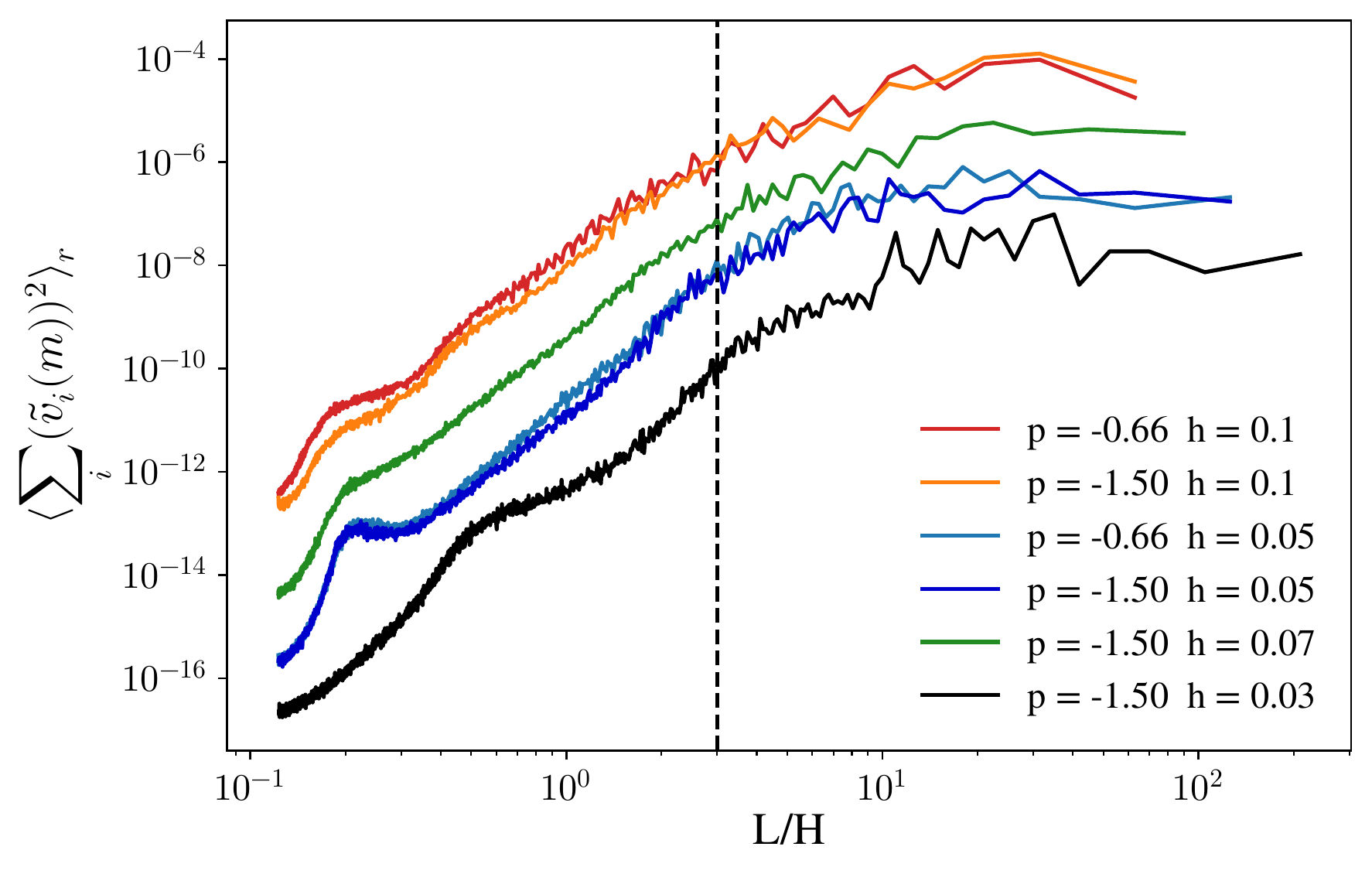}
    \caption{Specific kinetic energy spectrum as a function of turbulent length-scale in disk scale heights. The dashed line marks the scale $L=3H$, in close proximity of which the spectrum of all simulations shows a break in the power-law.}
    \label{fig:ekin_L}
\end{figure}

In figure \ref{fig:ekin_L} we plot the same spectrum as a function of the turbulent length scale $L = 1/k$ associated with the wavenumber m in units of the pressure scale height $H$. We find that the spectrum flattens to a shallower power law for all cases at or around $L = 3 H$, indicating a universal energy injection scale.\review{ We also observe that the energy is deposited at scales ranging from $L\approx 10 - 50\,H$ in all simulations, explaining the shift in preferred wavenumber of the RWI at different $h/r$.}

\section{Vortex formation and structure: Dependence on disk conditions}
\label{sec:results2}
To identify vortices in our simulations, we use the vertical component of the vorticity, defined as the rotation of the velocity vector $\vec{v}$:
\begin{equation}
    \omega_z = \left(\nabla \times \vec{v}\right)_z \qquad.
\end{equation}
In this section, we use the vorticity scaled by the orbital frequency $\Omega$ to identify and characterise the vortices forming in the disks at different values of p and h.

\subsection{Midplane vorticity}
\begin{figure}
    \centering
    \includegraphics[width=\columnwidth]{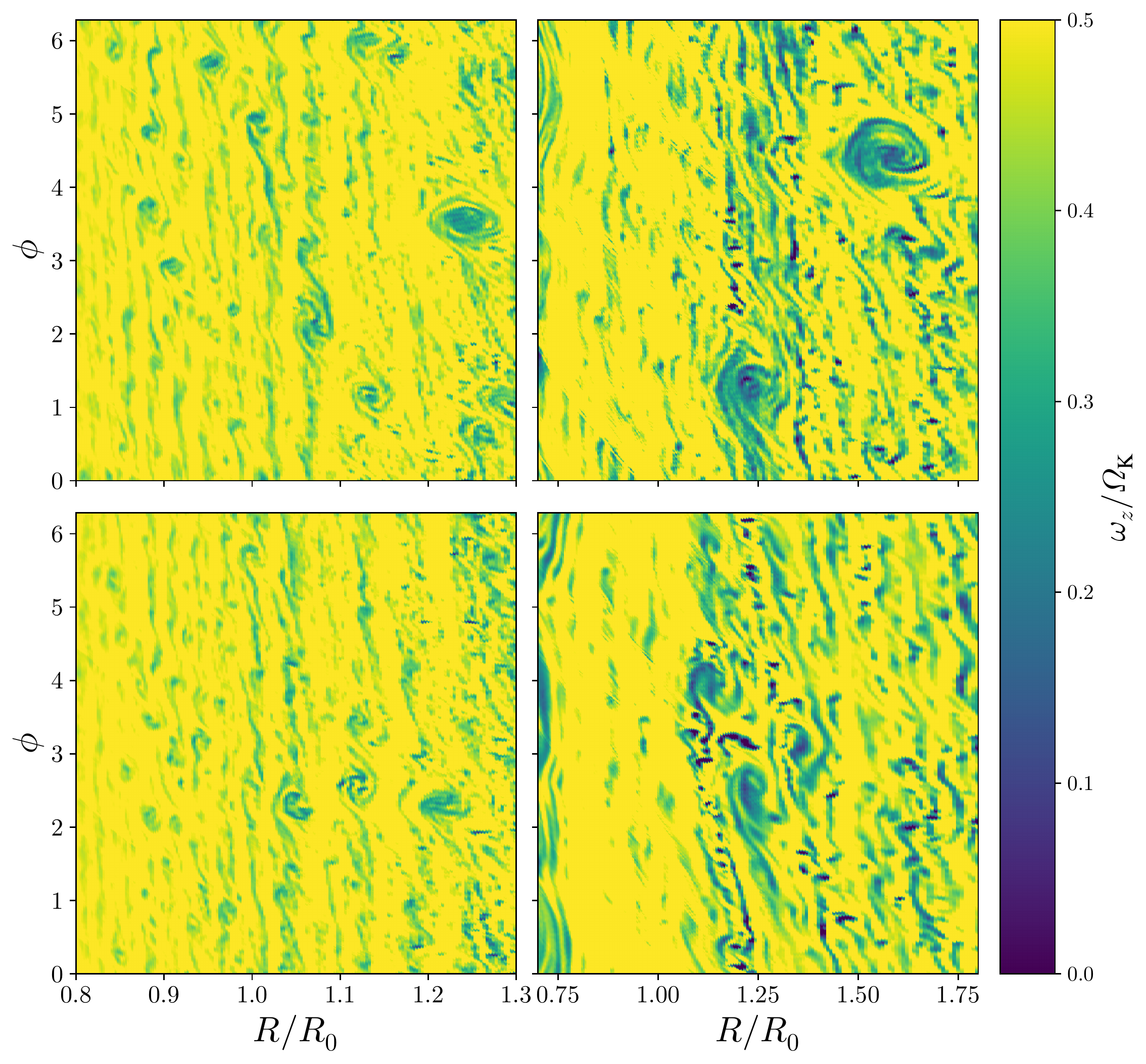}
    \caption{Midplane vorticity for different values of p and h. The top row shows simulations with $p=-0.66$ and the bottom row with $p=-1.5$. For the left column, the aspect ratio of the simulations is $h=0.05$, resulting in less extended structures than seen in the right column where $h=0.1$ .}
    \label{fig:vortpH}
\end{figure}

\begin{figure}
    \centering
    \includegraphics[width=\columnwidth]{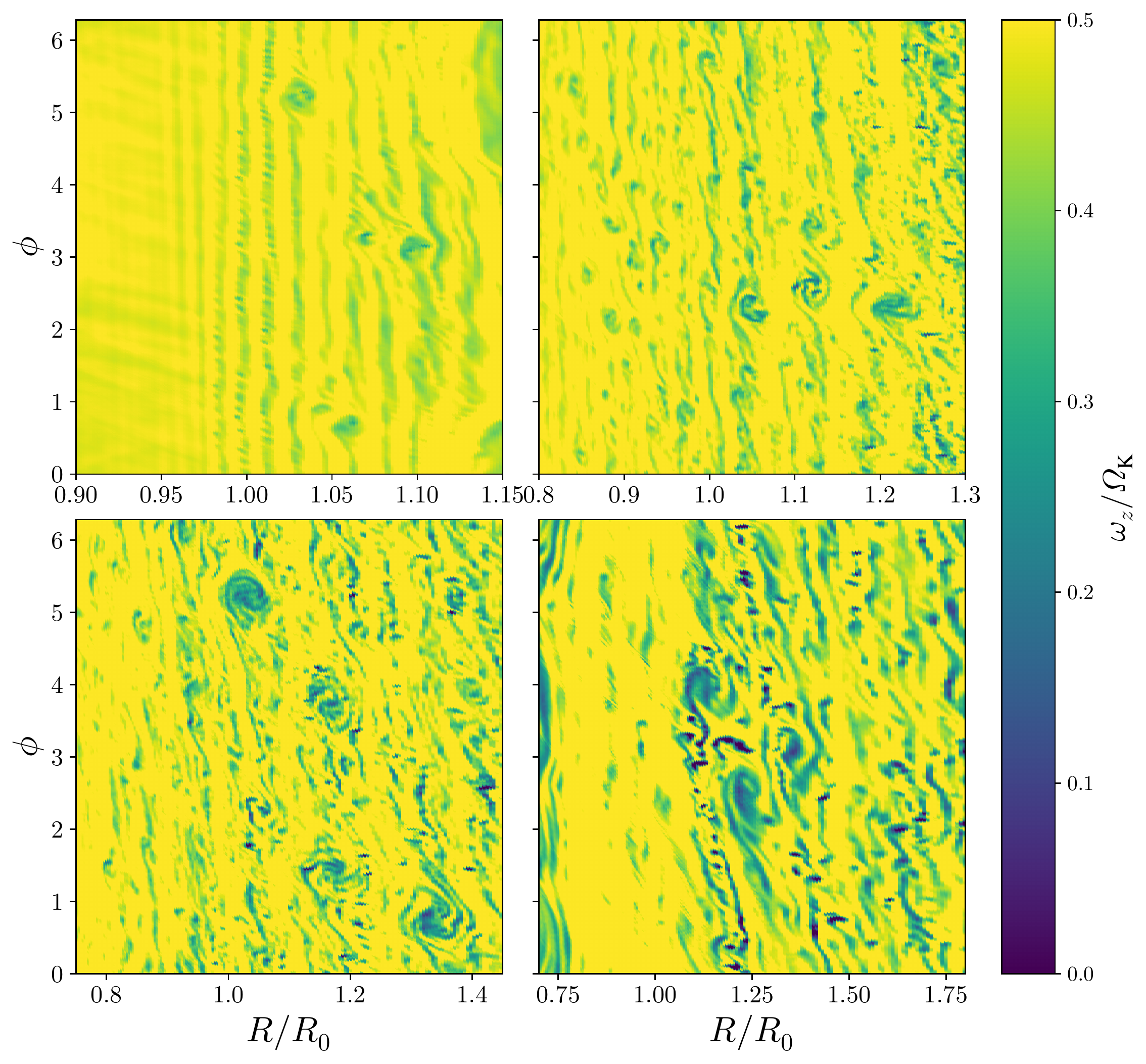}
    \caption{Midplane vorticity for different values of the disk aspect ratio $h$. The top left panel shows the simulation with the smallest h=0.03, increasing to $h=0.05$ in the top right, $h=0.07$ in the bottom left and $h=0.1$ in the bottom right panel. All simulations have initially $p=-1.5$. Note the increase in vorticity and decrease of overall structure with increasing $h$. }
    \label{fig:vortH}
\end{figure}

Figures \ref{fig:vortpH} and \ref{fig:vortH} show the midplane value of the vertical vorticity for the simulations with different 
$p$ and $h$ after 900 reference orbits. We find large vortices forming in all our simulations irrespective of the values assumed for $p$ and $h$. Each simulation has formed between 2 and around 10 vortices at this point in the simulation, but there is no correlation of the number of vortices with the chosen parameters.
The strength of the vortices however shows a correlation with the disk aspect ratio, best seen in figure \ref{fig:vortH}. The vortices in the bottom left panel corresponding to the simulation p1.5h0.1 appear to have lower absolute vorticity than the simulations in the top row corresponding to simulations p1.5h0.03 and p1.5h0.05. Because the disk has overall vorticity $\omega_z = \frac{1}{2} \Omega_\mathrm{K}$, the lower absolute vorticity in the centre of the vortices in the lower right panel corresponds to a larger relative vorticity and therefore vortex strength.

In figure \ref{fig:vortH} it can also be seen that in both simulations in the bottom row two or more vortices are in close radial proximity to each other. Therefore the vortices influence each other and each will eventually merge into one vortex. 

We also find that the run p1.5h0.03 (top left panel) shows a ordered band structure in vorticity, which does not appear in the other simulations at this stage, although the simulations with $h=0.05$ show axisymmetric bands at some radii. The band structure however appears for all simulations during the growth phase of the VSI, but it breaks down soon after. We discuss the implications of this in section \ref{sec:disc}.

In Appendix A we show the influence of the azimuthal resolution on the vorticity. For a resolution of $n_\phi \geq 256$ we find large vortices forming, corresponding to a minimal resolution of 4 cells per 
$H$ needed to resolve the non-axisymmetries of the non-linear stage of the VSI and the generation of the vortices through the Rossby-Wave-Instability mechanism as discussed in section \ref{sec:disc}. 

\subsection{Vortex size}

\begin{figure*}
    \centering
    \includegraphics[width=0.8\linewidth]{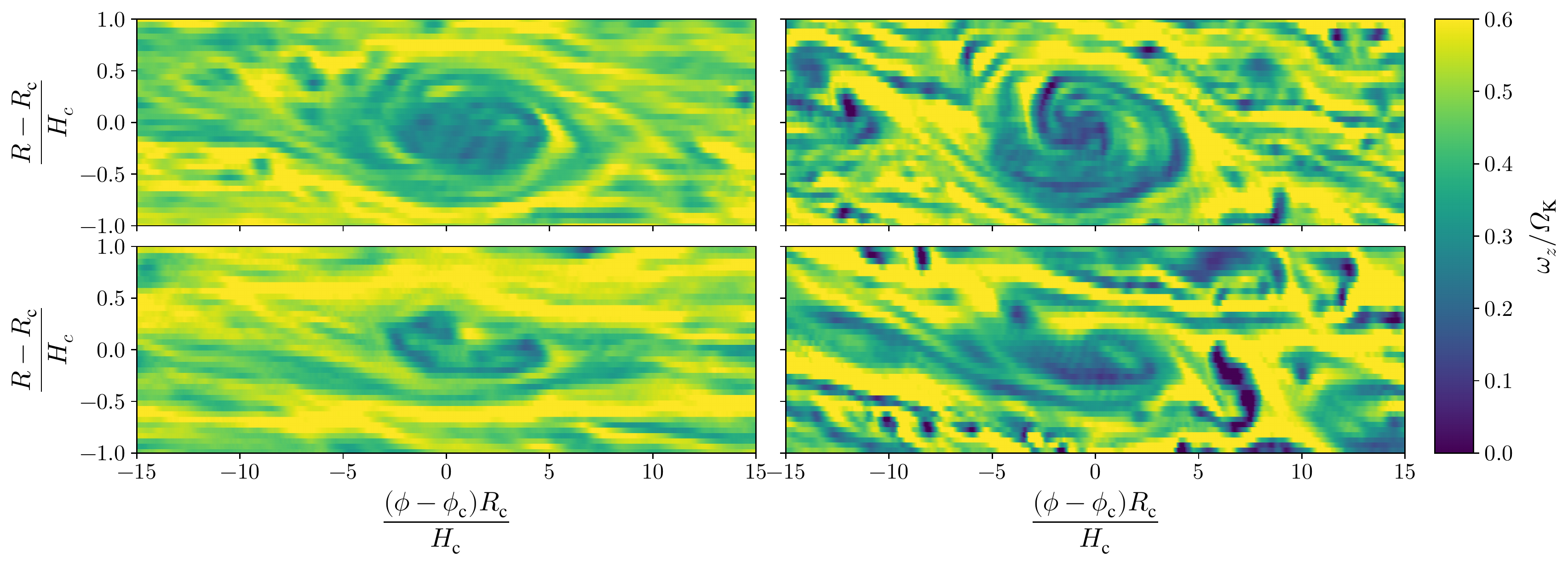}
    \caption{Vortex extent for different values of $p$ and $h$.
    The coordinate systems are positioned at the centres of the vortices,
    $R_c, \phi_c$. The simulations and the respective centre coordinates used are: p0.66h0.05 with $R_c,\phi_c = 1.25, 3.5$ (top left); p0.66h0.1 with $R_c,\phi_c = 1.6, 4.5 $ (top right); p1.5h0.05 with $R_c,\phi_c = 1.05, 2.25$ (lower left) and p1.5h0.1 with $R_c,\phi_c = 1.25,2.5$ (lower right).}
    \label{fig:vortZoompH}
\end{figure*}

\begin{figure*}
    \centering
    \includegraphics[width=0.8\linewidth]{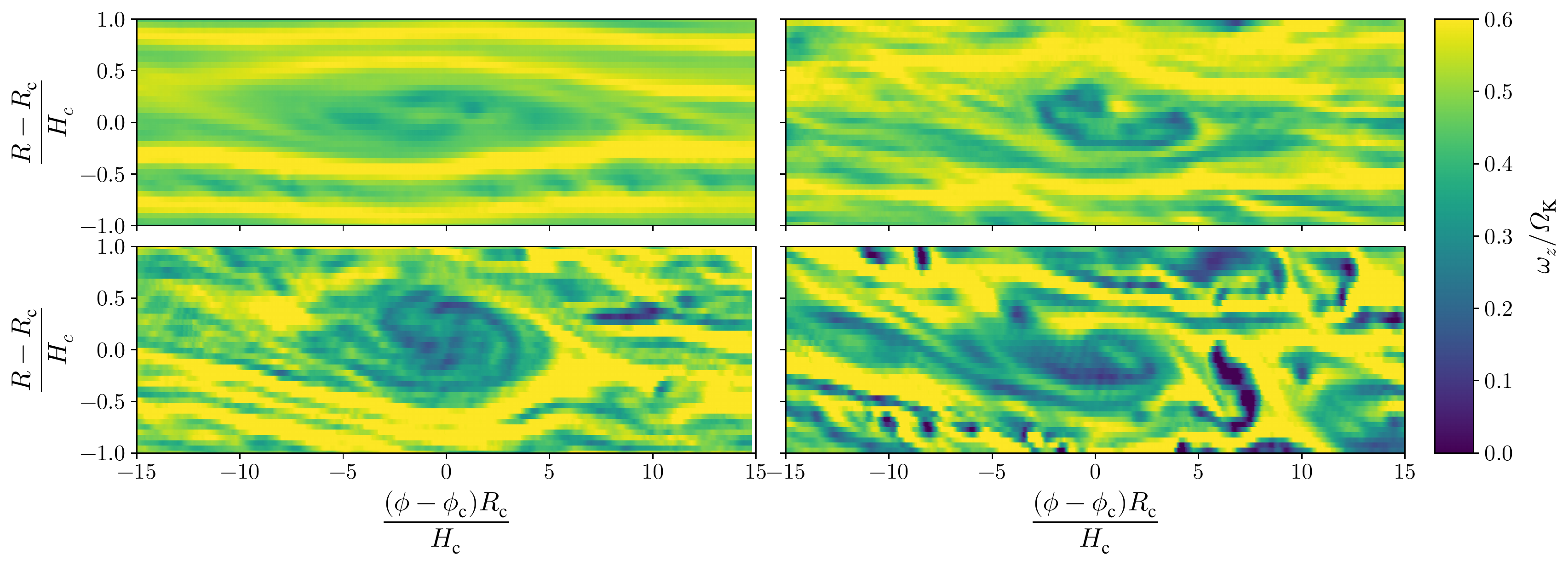}
    \caption{Vortex extent for different values of h and $p=-1.5$. The simulations shown, and the respective centre coordinates used, are: p1.5h0.03 with $R_c,\phi_c = 1.03, 5.25 $ (top left); p1.5h0.05 with $R_c,\phi_c = 1.05, 2.25$ (top right); p1.5h0.07 with $R_c,\phi_c = 1.03, 5.25$ (lower left) and p1.5h0.1 with $R_c,\phi_c = 1.25, 2.5$ (lower right).}
    \label{fig:vortZoomH}
\end{figure*}

Because we are also interested in the size and shape of the vortices formed, we choose one vortex from each simulation for a close up inspection. Figures \ref{fig:vortZoompH} and \ref{fig:vortZoomH} show the vortices taken from figures \ref{fig:vortpH} and \ref{fig:vortH}, respectively, in a local coordinate system centred on the vortex. We express the radial and azimuthal coordinate as a function of the local pressure scale height for easier comparison. The centre coordinates of each panel are listed in the corresponding figure caption.

We find that the vortices share a common size of 1-1.5 local scale heights in radial diameter and an azimuthal size between 10 and 20 $H$, leading to aspect ratios $\chi = r \Delta \phi / \Delta r$ in the range of 8.5 - 20.
Figure \ref{fig:vortZoomH} again shows the dependence of the relative vorticity on the disk scale height, but no correlation of vortex size or aspect ratio with disk aspect ratio h can be found. Figure \ref{fig:vortZoompH} seems to indicate that the vortices are larger for simulations with $p=-0.66$, but as the vortices depicted from the simulation p1.5h0.1 are currently merging, this could be artificial. This is supported by the fact that the top left panel of figure \ref{fig:vortpH} shows run p0.66h0.05 to also form vortices with a sizes comparable to the ones found in p1.5h0.05.

\subsection{Vortex evolution}
To track the radial position of the vortices over time, we use the same technique as employed in \citetalias{Manger+Klahr2018}. Therein, we calculate the radial positions of the vortices in each timestep by first applying a box filter to the vertical vorticity to eliminate all structures smaller than 1 $H$ in radius and 6 $H$ in azimuth. Then the azimuthal average of the vorticity is subtracted from this to exclude possible zonal flows and then the minimum value in azimuthal direction is calculated.

\begin{figure}
    \centering
    \includegraphics[width=\columnwidth]{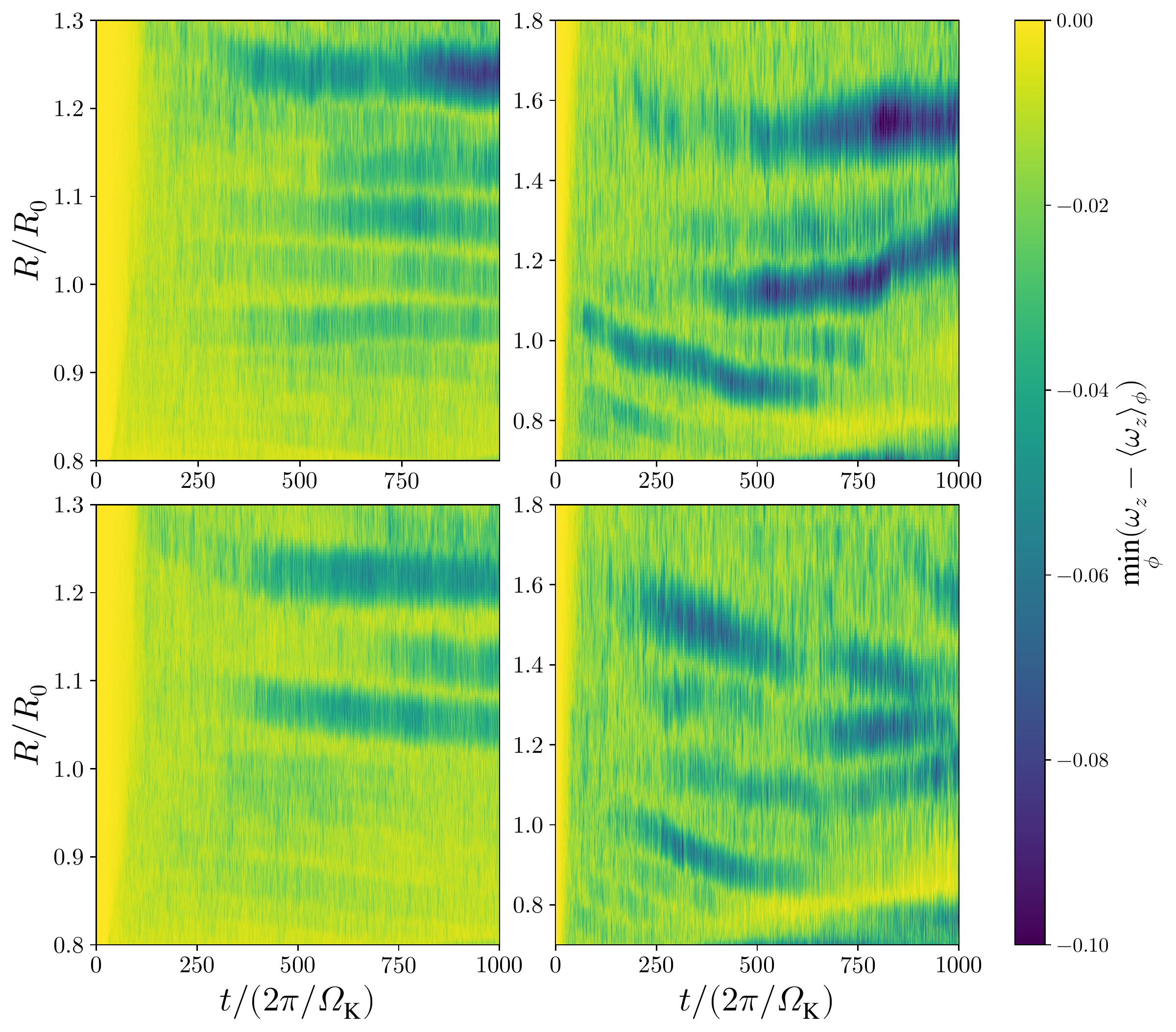}
    \caption{Evolution of the vortex radial position as a function of time. The panels are sorted as in figure \ref{fig:vortpH} with columns corresponding to aspect ratio and rows to density gradient. The colour shows the azimuthal minimum of the vertical vorticity subtracted by the azimuthally averaged vorticity. To extract only larger scale minima, we apply an image filter prior to the calculation.}
    \label{fig:vortRT_p-h}
\end{figure}

\begin{figure}
    \centering
    \includegraphics[width=\columnwidth]{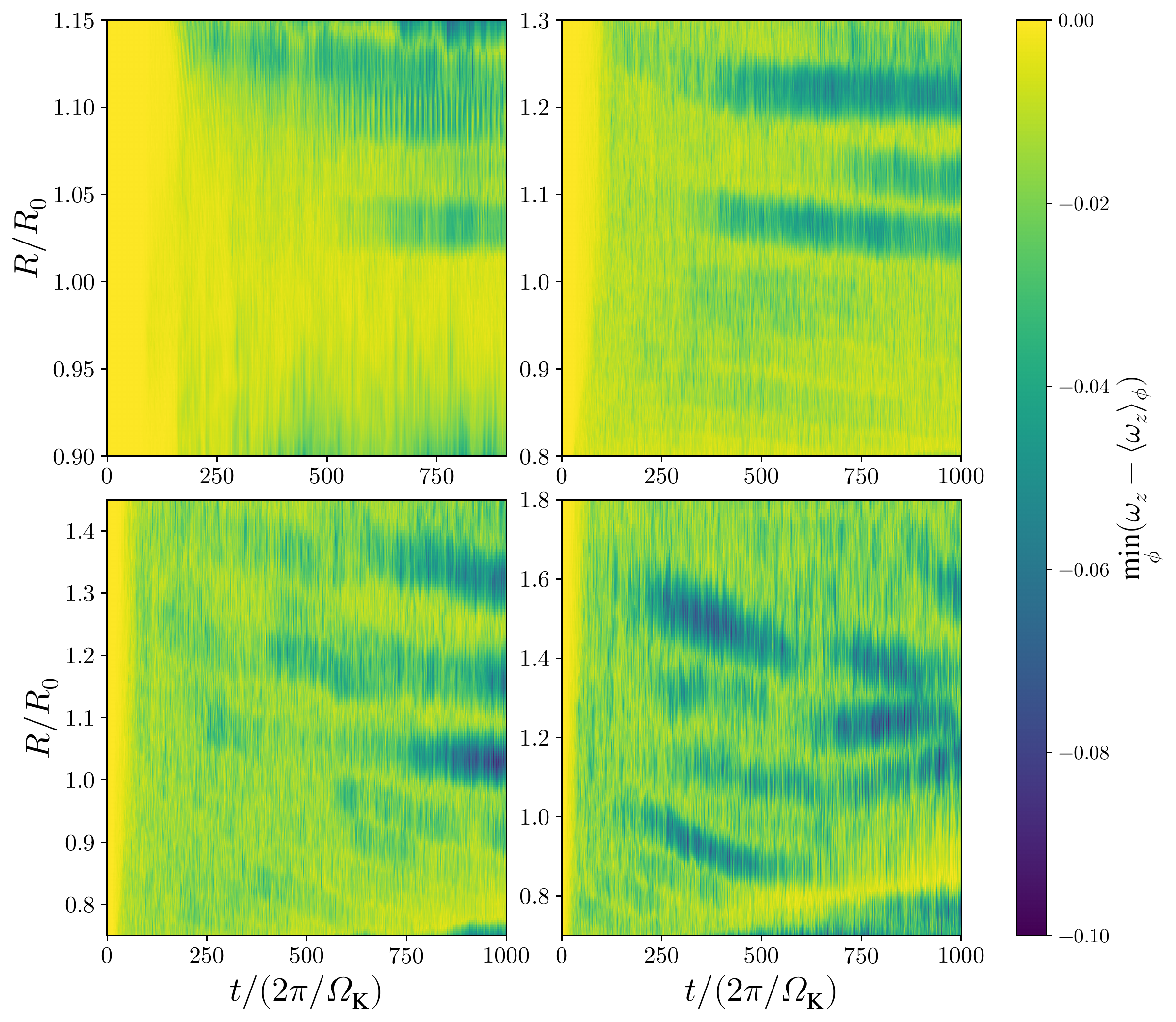}
    \caption{Same as figure \ref{fig:vortRT_p-h}, but now for the different disk aspect ratios as in figure \ref{fig:vortH}.}
    \label{fig:vortRT_h}
\end{figure}

The results are shown in figures \ref{fig:vortRT_p-h} and \ref{fig:vortRT_h}. In all cases, we find the formation of long lived vortices at the radial positions of the vortices depicted in figures \ref{fig:vortpH} and \ref{fig:vortH},
 though the time after which the stable vortices appear seems not directly correlated to neither density gradient nor aspect ratio. For example, the simulations p1.5h0.05 and p0.66h0.1 show stable vortex formation early on, while p1.5.h0.1 first shows intermittent large vortices before forming long time stable vortices. The same can be found upon inspecting additional frames of the run p1.5h0.07 which in figure \ref{fig:vortRT_h} only shows strong stable vortices emerging well after 750 orbits, but vortices can be found much earlier in the simulation, though their constant interaction makes it hard to detect them with this method.
 An exception in both the time of onset of vortex formation and the number of vortices formed is again run p1.5h0.03 which shows only one stable vortex forming after around 600 orbits, though this could be again attributed to the longer evolution timescale of the VSI itself.
 
 Upon inspection of the time series of the midplane vorticity, we also find that even if a vortex has established itself, new vortices can be formed at the same radial distance to the star, of which one example can be seen in the bottom left panel of figure \ref{fig:vortH}. The newly formed vortex catches up with the dominant vortex after a few tens to hundred orbits and is eventually absorbed into the stronger vortex. We also find evidence of this in other runs, e.g. p0.66h1.5, but as both examples occur for vortices located close to the reference radius we cannot exclude that this happens for other vortices also. The effect is most easily observed close to $R_0$ because we take one snapshot after each completed orbit at $R_0$, leading the azimuthal position of the vortices with $R> R_0$ and $R<R_0$ to drift due to the radial dependence of $\Omega$.
 
 Different to our results from \citetalias{Manger+Klahr2018}, we observe vortex destruction in our simulations with $h=0.1$. In both simulations, vortices formed early and close in in the disk are destroyed after ca. 500 and 300 orbits for $p=0.66$ and $p=-1.5$, respectively. For the case p=-0.66 this is due to a vortex forming radially close to the vortex and interacting with it until it is destroyed. For the case $p=-1.5$ something similar seems to happen, but we cannot exclude the influence of the boundary in this case.

We also observe the migration of the vortices over time in all cases except p0.66h0.05 and p1.5h0.03. For the former this is likely due to the large amount of vortices formed in the disk which prevents radial migration due to the interaction of the vortices with each other. For the latter, it is simply the late time of the formation of the vortices that prohibits us from detecting migration, though it is possible that it occurs later on. For the migrating cases, the migration is stronger for simulations with larger aspect ratio. This can be explained by the fact that the vortices formed in disks with larger h are stronger, but we cannot exclude the influence of the disk surface density gradient, which changes most for the cases with large h due to overall mass loss.

\section{Discussion}
\label{sec:disc}
In this section, we discuss our results presented \ref{sec:results1} and compare them to other models. We start with the  turbulent velocities and then present an interpretation of the relationship between the stress-to-pressure ratio $\alpha$ and $h$.

\subsection{Turbulence growth rates}
In section \ref{sec:results1}, we showed the growth rate of the VSI in our simulations with $h=0.1$ to be $\Gamma \approx 0.4$ and $\Gamma=0.2$ for the simulations with $h=0.05$. These values of $\Gamma$ are in good agreement with the values obtained in \citetalias{Manger+Klahr2018} and \citet{Stoll+Kley2014}, respectively. The observed linear scaling of most of the reported values with aspect ratio is expected by linear theory. \citet{Nelson+2013} showed that for nearly isothermal disks the growth rate can be to first order expressed as
\begin{equation}
\Gamma \sim \lvert q \rvert h \Omega \qquad .\label{eqn:VSIgrowth}
\end{equation}
 
To compare our results with the theoretical ones presented in \citet{Lin+Youdin2015} we calculate the h independent growth rate 
\begin{equation}
    \sigma = \frac{\Gamma}{h\Omega} \qquad .
\end{equation}
For the simulations with $h=0.1,\,0.07$ and 0.05 we obtain  $\sigma=0.64$, which is in good agreement with the results presented by \citet{Lin+Youdin2015} for a disk simulation with $z_\mathrm{max}=3H$ and radial wavenumber k=30. Because we obtained a smaller than expected growth rate for $h=0.03$, this case also has a lower $\sigma$ and does not align well with their largest expected growth rate.

We additionally observed that the onset of VSI growth is increasingly delayed with decreasing aspect ratio. The occurrence of this trend is consistent with the results of \citet{Nelson+2013}, who observed this behaviour for the kinetic energy in their 2D axisymmetric simulations.

\subsection{Stress-to-pressure ratio scaling law}

Our simulations with $h=0.1$ show a saturated $\alpha$ of $9\cdot 10^{-4}$. This value compares well with the values we obtained in \citetalias{Manger+Klahr2018}. It is however about one order of magnitude larger than the values reported in both \citet{Flock+2017} and \citet{Flock+2020}, who reach a similar scale height in their radiation hydrodynamics simulation. This is likely due to the cooling time varying between $\tau_\mathrm{cool} = 10^{-2} - 10^{-4}$ orbital periods in their simulations in contrast to our fixed cooling time of approximately $10^{-4}$ orbits.
An investigation of the influence of the cooling time on the VSI strength is however beyond the scope of this work. For our simulations with $h=0.05$ we find a saturated $\alpha$ of $10^{-4}$, which is comparable to the values of a few times $10^{-4}$ reported for the high resolution cases in \citet{Stoll+Kley2014}. 

\begin{figure}
    \centering
    \includegraphics[width=\columnwidth]{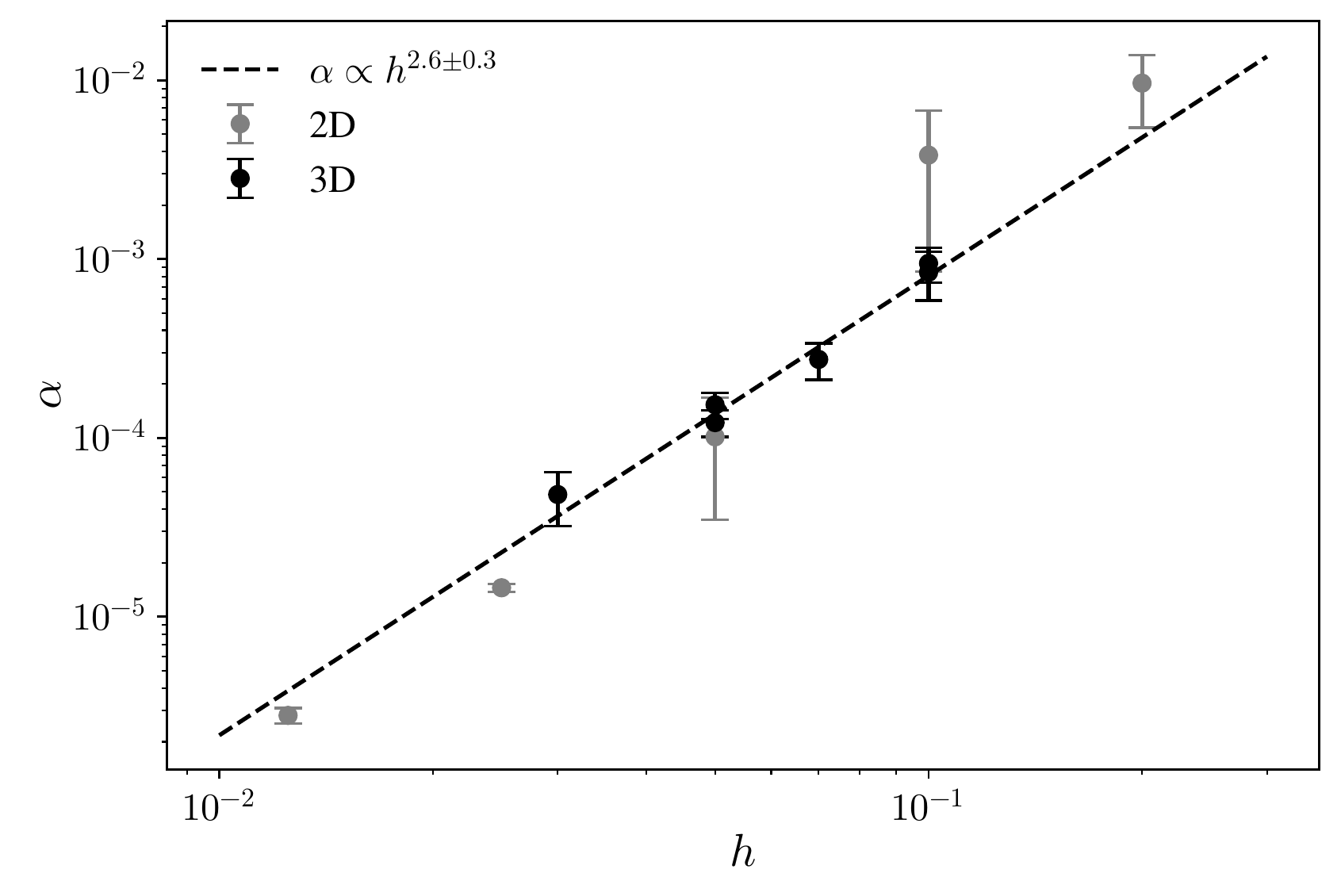}
    \caption{Time and space averaged $\alpha$-values as a function of disk scale height. Clearly visible is the power law relationship between both, which we determine to be $\alpha \propto h^{2.6}$ for the 3D simulations. Although $\alpha$ is not strictly defined in 2D, we find good agreement with the 3D data using a time averaged to determine $\langle v_\phi \rangle$.}
    \label{fig:alphavsh}
\end{figure}

\begin{figure}
    \centering
    \includegraphics[width=\columnwidth]{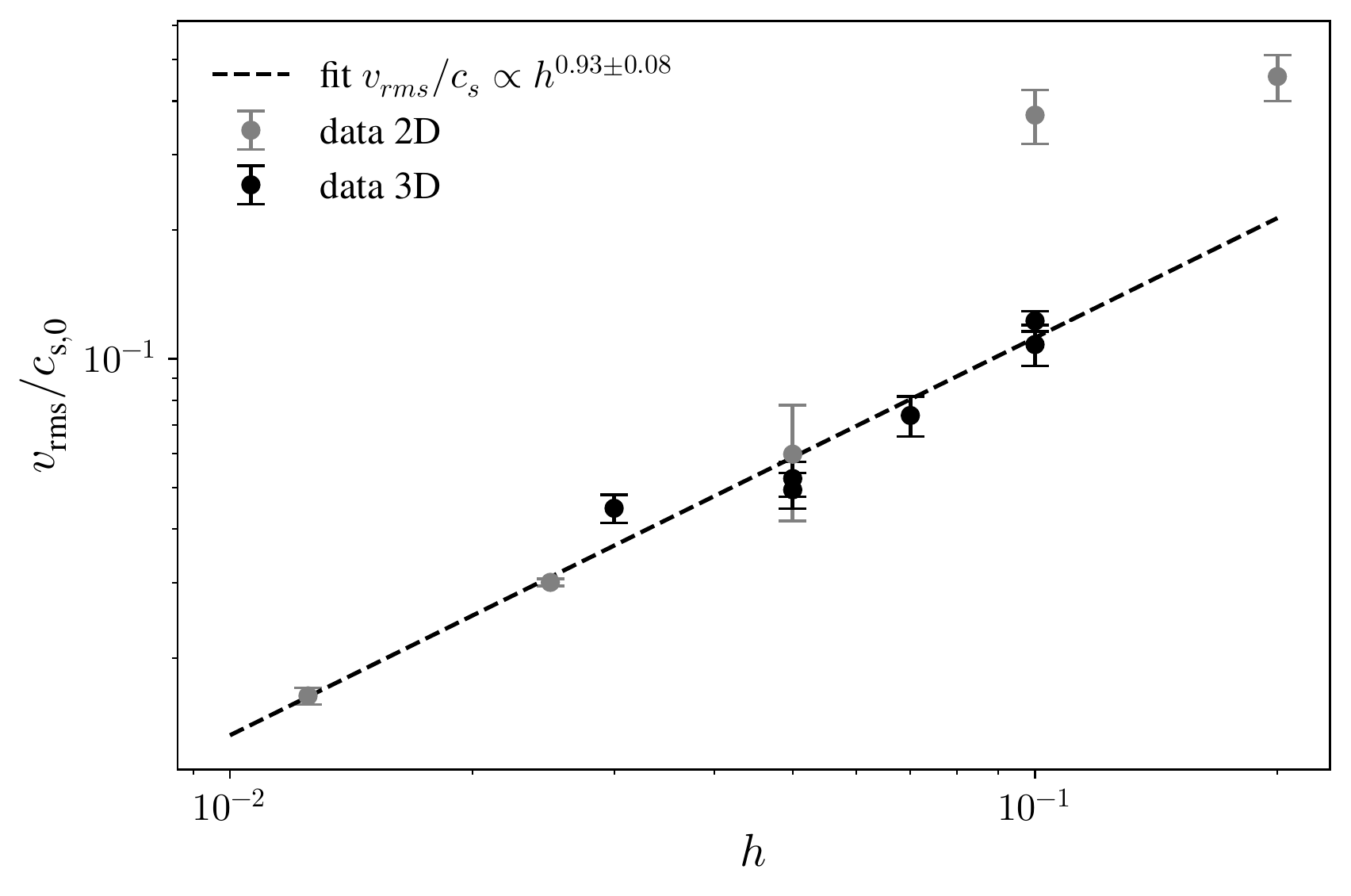}
    \caption{Time and space averaged $v_\mathrm{rms}$-values as a function of disk scale height. Clearly visible is the linear relationship between both. The outliers at high $h$ can be explained by the non-linear state of the VSI in 2D.}
    \label{fig:vrmsvsh}
\end{figure}

\review{
When we plot the $\alpha$ values in our 3D simulations as a function of the pressure scale height $h$  (see Fig.\ref{fig:alphavsh}) we find a scaling of 2.6, which seeks for an explanation. For that purpose we measure the strength of the VSI, i.e.\ the r.m.s. velocity of the gas as a function of scale height $h$ (see Fig.\ref{fig:vrmsvsh}) and find it to scale linearly, which is easy to explain.}

\review{
The VSI is driven by vertical shear $S = |\partial_z v_\phi| = R |\partial_z \Omega|$ (see Eq. \ref{eqn:Omega}) over a scale height $H$:
\begin{equation}
S \cdot H = R \Omega_K \frac{1}{2} \left(\frac{H}{R}\right)^2 q = c_s \frac{q}{2} h,
\end{equation}
thus if $v_{\rm r.m.s.} \propto S H$, then we find:
\begin{equation}
v_{\rm r.m.s.} \propto c_s \frac{q}{2} h,
\end{equation}
exactly what we find in Fig. \ref{fig:vrmsvsh}. We ran additional 2D simulations (listed in table \ref{tab:simParam2d}) to confirm this trend over an even wider range of $h$. Although $\alpha$ as defined in equation \ref{eq:alpha} is not strictly applicable to 2D simulations, we used a time average to determine the steady state $v_\phi$ velocity component  and the results align well to the linear scaling law retrieved from the 3D data for the simulations with $h<0.1$. The higher values of $v_\mathrm{rms}$ and $\alpha$ for the 2D simulations with $h>0.1$ can be explained by the non-linear state of the 2D simulations developing large scale motions not present in 3D.}

\review{
As now $\alpha$ measures the Reynolds stresses scaled with the speed of sound, we find
\begin{equation}
\alpha \propto \frac{v_{\rm r.m.s.}^2}{c_s^2} \propto h^2,
\end{equation}
clearly leading to a steep rise in $\alpha$ as function of $h$.
This estimated scaling is not as steep as the one we measured in Fig.\ref{fig:alphavsh}, which probably needs additional effects. One suggestion would be the decreased pitch angle of the spiral density waves that actually transport the angular momentum \citep{Rafikov2002} with $h$. Thus $h$ modifies the correlation between radial $v_r$ and azimuthal $ v_\phi$ velocity fluctuations. In other words, the fewer pressure scale heights fit into the circumference of the disk, the less tightly the spirals are wrapped and the stronger the angular momentum transport becomes. 
}

\begin{table}
\renewcommand\arraystretch{1.3}
\centering
\begin{tabular}{l c c l c}
\hline
model & $r_\mathrm{in, out}/R_0$& $\mathrm{N}_\mathrm{r}\times\mathrm{N}_\theta \times \mathrm{N}_\varphi$ & $h$ & $\langle \alpha \rangle /10^{-4}$ \\ \hline
p1.5h0.01  & $0.96-1.03$& 512$\times$512$\times$1 & $0.0125$ & $0.028\pm0.003 $\\
p1.5h0.02 & $0.92-1.07$& 512$\times$512$\times$1 & $0.025$ & $0.145\pm0.007 $\\
p1.5h0.05 & $0.85-1.15$ & 512$\times$512$\times$1 & $0.05$ & $1.0\pm0.7  $\\
p1.5h0.1 & $0.7- 1.3$ & 512$\times$512$\times$1 & $0.1$ & 38$\pm30  $\\
p1.5h0.2 & $0.6- 1.4$ & 512$\times$512$\times$1 & $0.2$ & $96\pm42  $\\
\hline
\end{tabular}
\caption{List of additional 2D simulation parameters. From left to right: model name, radial domain size, numerical grid size, disk aspect ratio and space and time averaged stress to pressure value.  The vertical and azimuthal domain sizes for all simulations are $\theta = \pm 3.5 H$ and $\phi = 0 - 2\pi$, respectively, and we assume $p=-1.5$ throughout.
}
\label{tab:simParam2d}
\end{table}


\subsection{Vortex formation and evolution}
In \citetalias{Manger+Klahr2018} we put forward the hypothesis that the Rossby-Wave-Instability (RWI) is working as a secondary instability in our simulations. \citep{Lovelace+1999} showed that the RWI develops in disks that form regions with an extremum in $\mathscr{L}$, defined as
\begin{equation}
\mathscr{L} = \frac{\Sigma}{2\omega_z}\,\left(\frac{P}{\Sigma^\gamma}\right)^{2/\gamma} \qquad. \label{eqn:RWI}
\end{equation}
Within a region around this extremum, standing Rossby waves develop and amplify. Eventually multiple vortices form within this region which subsequently merge into one large vortex.\review{ Looking at $\mathscr{L}$ as a function of radius supports this: Figure \ref{fig:RossbyL} shows $\mathscr{L}$ after 200 and 500 orbits, where the profile of $\mathscr{L}$ changes significantly in amplitude for the cases with $h=0.05$ and $0.1$ where the RWI has formed vortices well before 500 orbits, whereas in the remaining runs RWI is still developing, and therefore $\mathscr{L}$ retains its large amplitude. For those cases, it is more instructing to look at $\mathscr{L}$ after 500 and 700 orbits, respectively. There we can see that for the case with $h=0.03$ the amplitude of $\mathscr{L}$ at the position of the vortex is declining between the two times, and the vortex appears after around 600 orbits. This finding is in line with \citet{Meheut+2010}, who showed that the extremum in $\mathscr{L}$ diminishes after the instability saturates. For the case with $h=0.07$ the vortices form very late in the simulation, so the extremum grows as time progresses, and we see a strong extremum only at time $t=700$ orbits, which will eventually develop into a vortex about 50-100 orbits later.}

\review{Additionally, we find small vortices forming in all simulations with $h \geq 0.05$, which can be seen in figures \ref{fig:vortZoompH} and \ref{fig:vortZoomH}. These smaller vortices have been previously identified in \citet{Richard+2016} and are in line with the work of \citet{Latter+Papaloizou2018}, who showed that a parasitic Kelvin-Helmholtz instability (KHI) develops off the VSI modes, leading to the formation of small vortices. We however argue that these KHI-vortices are transient and that the instability is not responsible for the larger long-lived vortices we find in our simulations.}

In our simulations with $h=0.03$ and $h=0.05$ presented in section \ref{sec:results2}, we also find azimuthal bands of low vorticity in addition to the vortices forming in the disk, generated by the vertical structure of the body modes of the VSI. 
We interpret these azimuthal bands as zonal flow like structures forming from the VSI body modes in the disk, representing the axisymmetrical nature of the VSI. If these modes indeed act as zonal flows, they likely still exist also in the simulations with large $h$ as an underlying construct obscured by the high turbulence level present. This would allow for the formation of dust rings in the presence of e.g. magnetic fields inhibiting the long term survival of the vortices. 

\review{The vortices shown in figures \ref{fig:vortZoompH} and \ref{fig:vortZoomH} also show internal turbulent structure. As our simulations are conducted in 3D, the vortices are in principle susceptible to the Elliptic Instability \citep[EI,][]{Lesur+Papaloizou2009}. But the EI is easily suppressed by numerical diffusion, especially for vortices with $\chi \gtrsim 10$. We therefore argue that the EI is not significantly effecting the vortices in our simulations. The intrinsic turbulent structure of the vortices can more easily be explained with the intrinsic turbulence generated by the VSI and the small Kelvin-Helmholtz parasite vortices generated by it \citep{Latter+Papaloizou2018}.}

\section{Conclusions}
\label{sec:conlc}
In this work we present the first 3D high resolution parameter study on the Vertical Shear Instability. We focus our attention on how the initial density slope $p$ of the disk and the disk aspect ratio $h$ impacts the VSI and vortex formation. The dependence on the cooling time and temperature gradient will be the the focus of future work. For now, we have assumed $\tau \approx 10^{-4}$ and $q=-1$.

We find that the VSI is capable to support angular momentum transport with $\alpha$ values up to  \review{$\alpha = 10^{-3}$} for the largest scale height of \review{$h = 0.1$}.  The $\alpha$ values we find in our simulations are comparable but slightly lower than the values we reported in \citetalias{Manger+Klahr2018} and the values found by \cite{Stoll+Kley2014} for the respective $h$ used. The values reported differ however from values obtained from radiation hydrodynamics simulations of the VSI \citep{Flock+2017,Flock+2020}, which will need to be be investigated in the future.

\review{With decreasing scale height, the $\alpha$ value decreases down to a value of a few times $10^{-5}$ for the lowest scale height investigated. Overall, we find $\alpha$ to scale with $h$ roughly as $h^{2.6}$. We explain this behaviour with the driving shear of the VSI, which increases as $h$, generating a dependence of $\alpha \propto h^2$. The additional $h$ dependence can be attributed to the formation of spiral density waves, which also transport angular momentum \citep{Rafikov2002}. The dependence of $\alpha$ on $h$ could have a profound influence on the disk structure which should be investigated further.}

The growth rates of the VSI we find are in good agreement with previous studies of \citetalias{Manger+Klahr2018} and \cite{Stoll+Kley2014}. In the vertical direction, the quantitative behaviour we observed in \citetalias{Manger+Klahr2018} is recovered. 
Contrary to the aspect ratio of the disk, the density gradient does not have any influence on the disk turbulence. This result is expected, as the VSI itself is not sensitive to the density gradient.

We find that the VSI is able to seed multiple vortices in all considered parameter combinations and the vortices in all simulations live for hundreds of orbits, reinforcing the findings of \citetalias{Manger+Klahr2018}. The vortices generally have a radial diameter between 1 and 1.5 local scale heights, which is close to the diameter of $2 H$ allowed by the disks radial shear. Most vortices have aspect ratios of $\chi \simeq 8$ but values of up to 20 are found. The time at which the first stable vortex appears is not correlated with any of the investigated parameters, and it is likely that such a time is random. We also do not find a correlation between the number of vortices found simultaneously in the disk and any of the investigated parameters.
We also observe that the turbulence sustained by the VSI steadily creates new vortices even at radii at which a large scale vortex has already been established for a longer period of time. This could indicate that the VSI constantly works to replenish the vorticity gradient that drives the Rossby-Wave-Instability, which once replenished sufficiently creates new vortices. These new, weaker vortices are then eventually absorbed by the larger vortex already present.

In this work, we considered the disk to be comprised purely of gas. In reality, the disk contains about 2\% solids, which have been shown to change the buoyancy of the disk and can therefore have an impact on the growth of the VSI especially near the midplane \citep{Lin2019,Schaefer+2020}. Future simulations should therefore investigate whether vortices also emerge in a dusty disk when the VSI is suppressed close to the midplane.\review{ We also neglect the influence of magnetic fields in this work. \citet{Cui+Bai2020} showed that the VSI can in principle coexist with magnetically launched winds at the surface, but future work should address whether the non-axisymmetric effects found in this work are affected by non-ideal MHD processes.}

\section*{Acknowledgements}
This research has been supported by the Deutsche Forschungsgemeinschaft Schwerpunktprogramm (DFG SPP) 1833 "Building a Habitable Earth" under contract KL1469/13-1 "Der Ursprung des Baumaterials der Erde: Woher stammen die Planetesimale und die Pebbles? Numerische Modellierung der Akkretionsphase der Erde."
WK and HK acknowledge funding by the DFG Research Unit 'Transition Disks' under grant KL 650/29-1. MF acknowledges financial support from the
European Research Council (ERC) under the European Union's Horizon 2020
research and innovation programme (grant agreement No. 757957).
The authors gratefully acknowledge the Gauss Centre for Supercomputing (GCS) for providing computing time for a GCS Large-Scale Project on the GCS Supercomputer HAZEL HEN at H\"ochstleistungsrechenzentrum Stuttgart (www.hlrs.de). GCS is the alliance of the three national supercomputing centres HLRS (Universit\"at Stuttgart), JSC (Forschungszentrum J\"ulich), and LRZ (Bayerische Akademie der Wissenschaften), funded by the German Federal Ministry of Education and Research (BMBF) and the German State Ministries for Research of Baden-W\"urttemberg (MWK), Bayern (StMWFK) and Nordrhein-Westfalen (MIWF). Additional simulations were performed on the ISAAC cluster owned by the MPIA and the HYDRA cluster of the Max-Planck-Society, hosted at the Max-Planck Computing and Data Facility in Garching (Germany).

\bibliography{references,wkrefs}

\begin{thebibliography}{}
\makeatletter
\relax
\def\mn@urlcharsother{\let\do\@makeother \do\$\do\&\do\#\do\^\do\_\do\%\do\~}
\def\mn@doi{\begingroup\mn@urlcharsother \@ifnextchar [ {\mn@doi@}
  {\mn@doi@[]}}
\def\mn@doi@[#1]#2{\def\@tempa{#1}\ifx\@tempa\@empty \href
  {http://dx.doi.org/#2} {doi:#2}\else \href {http://dx.doi.org/#2} {#1}\fi
  \endgroup}
\def\mn@eprint#1#2{\mn@eprint@#1:#2::\@nil}
\def\mn@eprint@arXiv#1{\href {http://arxiv.org/abs/#1} {{\tt arXiv:#1}}}
\def\mn@eprint@dblp#1{\href {http://dblp.uni-trier.de/rec/bibtex/#1.xml}
  {dblp:#1}}
\def\mn@eprint@#1:#2:#3:#4\@nil{\def\@tempa {#1}\def\@tempb {#2}\def\@tempc
  {#3}\ifx \@tempc \@empty \let \@tempc \@tempb \let \@tempb \@tempa \fi \ifx
  \@tempb \@empty \def\@tempb {arXiv}\fi \@ifundefined
  {mn@eprint@\@tempb}{\@tempb:\@tempc}{\expandafter \expandafter \csname
  mn@eprint@\@tempb\endcsname \expandafter{\@tempc}}}

\bibitem[\protect\citeauthoryear{{Arlt} \& {Urpin}}{{Arlt} \&
  {Urpin}}{2004}]{Arlt+Urpin2004}
{Arlt} R.,  {Urpin} V.,  2004, \aap, 426, 755

\bibitem[\protect\citeauthoryear{{Balbus} \& {Hawley}}{{Balbus} \&
  {Hawley}}{1991}]{Balbus+Hawley1991}
{Balbus} S.~A.,  {Hawley} J.~F.,  1991, \mn@doi [Astrophysical Journal]
  {10.1086/170270}, 376, 214

\bibitem[\protect\citeauthoryear{{Barge} \& {Sommeria}}{{Barge} \&
  {Sommeria}}{1995}]{Barge+Sommeria1995}
{Barge} P.,  {Sommeria} J.,  1995, \aap, 295, L1

\bibitem[\protect\citeauthoryear{{Barranco}, {Pei}  \& {Marcus}}{{Barranco}
  et~al.}{2018}]{Barranco+2018}
{Barranco} J.~A.,  {Pei} S.,   {Marcus} P.~S.,  2018, \apj, 869, 127

\bibitem[\protect\citeauthoryear{{B{\'e}thune}, {Lesur}  \&
  {Ferreira}}{{B{\'e}thune} et~al.}{2017}]{2017A&A...600A..75B}
{B{\'e}thune} W.,  {Lesur} G.,   {Ferreira} J.,  2017, \mn@doi [\aap]
  {10.1051/0004-6361/201630056}, \href
  {https://ui.adsabs.harvard.edu/abs/2017A&A...600A..75B} {600, A75}

\bibitem[\protect\citeauthoryear{{Cui} \& {Bai}}{{Cui} \&
  {Bai}}{2020}]{Cui+Bai2020}
{Cui} C.,  {Bai} X.-N.,  2020, \mn@doi [\apj] {10.3847/1538-4357/ab7194}, 891,
  30

\bibitem[\protect\citeauthoryear{{Dzyurkevich}, {Flock}, {Turner}, {Klahr}  \&
  {Henning}}{{Dzyurkevich} et~al.}{2010}]{Dzyurkevich+2010}
{Dzyurkevich} N.,  {Flock} M.,  {Turner} N.~J.,  {Klahr} H.,   {Henning} T.,
  2010, \mn@doi [Astronomy and Astrophysics] {10.1051/0004-6361/200912834},
  515, A70

\bibitem[\protect\citeauthoryear{{Flock}, {Nelson}, {Turner}, {Bertrang},
  {Carrasco-Gonz{\'a}lez}, {Henning}, {Lyra}  \& {Teague}}{{Flock}
  et~al.}{2017}]{Flock+2017}
{Flock} M.,  {Nelson} R.~P.,  {Turner} N.~J.,  {Bertrang} G.~H.-M.,
  {Carrasco-Gonz{\'a}lez} C.,  {Henning} T.,  {Lyra} W.,   {Teague} R.,  2017,
  \apj, 850, 131

\bibitem[\protect\citeauthoryear{{Flock}, {Turner}, {Nelson}, {Lyra}, {Manger}
  \& {Klahr}}{{Flock} et~al.}{2020}]{Flock+2020}
{Flock} M.,  {Turner} N.~J.,  {Nelson} R.~P.,  {Lyra} W.,  {Manger} N.,
  {Klahr} H.,  2020, \mn@doi [\apj] {10.3847/1538-4357/ab9641}, 897, 155

\bibitem[\protect\citeauthoryear{{Fricke}}{{Fricke}}{1968}]{Fricke1968}
{Fricke} K.,  1968, Zeitschrift f{\"u}r Astrophysik, 68, 317

\bibitem[\protect\citeauthoryear{{Goldreich} \& {Schubert}}{{Goldreich} \&
  {Schubert}}{1967}]{Goldreich+Schubert1967}
{Goldreich} P.,  {Schubert} G.,  1967, \mn@doi [Astrophysical Journal]
  {10.1086/149360}, 150, 571

\bibitem[\protect\citeauthoryear{{Klahr} \& {Bodenheimer}}{{Klahr} \&
  {Bodenheimer}}{2003}]{Klahr+Bodenheimer2003}
{Klahr} H.~H.,  {Bodenheimer} P.,  2003, \apj, 582, 869

\bibitem[\protect\citeauthoryear{Klahr \& Hubbard}{Klahr \&
  Hubbard}{2014}]{Klahr+Hubbard2014}
Klahr H.,  Hubbard A.,  2014, \mn@doi [The Astrophysical Journal]
  {10.1088/0004-637X/788/1/21}, 788, 21

\bibitem[\protect\citeauthoryear{Kraichnan}{Kraichnan}{1971}]{Kraichnan1971}
Kraichnan R.~H.,  1971, Journal of Fluid Mechanics, 47, 525

\bibitem[\protect\citeauthoryear{{Latter} \& {Papaloizou}}{{Latter} \&
  {Papaloizou}}{2018}]{Latter+Papaloizou2018}
{Latter} H.~N.,  {Papaloizou} J.,  2018, \mnras, 474, 3110

\bibitem[\protect\citeauthoryear{{Lenz}, {Klahr}  \& {Birnstiel}}{{Lenz}
  et~al.}{2019}]{Lenz+2019}
{Lenz} C.~T.,  {Klahr} H.,   {Birnstiel} T.,  2019, \mn@doi [\apj]
  {10.3847/1538-4357/ab05d9}, 874, 36

\bibitem[\protect\citeauthoryear{{Lesur} \& {Papaloizou}}{{Lesur} \&
  {Papaloizou}}{2009}]{Lesur+Papaloizou2009}
{Lesur} G.,  {Papaloizou} J.~C.~B.,  2009, \aap, 498, 1

\bibitem[\protect\citeauthoryear{{Lesur} \& {Papaloizou}}{{Lesur} \&
  {Papaloizou}}{2010}]{Lesur+Papaloizou2010}
{Lesur} G.,  {Papaloizou} J.~C.~B.,  2010, \aap, 513, A60

\bibitem[\protect\citeauthoryear{{Li}, {Finn}, {Lovelace}  \& {Colgate}}{{Li}
  et~al.}{2000}]{Li+2000}
{Li} H.,  {Finn} J.~M.,  {Lovelace} R.~V.~E.,   {Colgate} S.~A.,  2000, {\apj},
  533, 1023

\bibitem[\protect\citeauthoryear{{Li}, {Colgate}, {Wendroff}  \& {Liska}}{{Li}
  et~al.}{2001}]{Li+2001}
{Li} H.,  {Colgate} S.~A.,  {Wendroff} B.,   {Liska} R.,  2001, {\apj}, 551,
  874

\bibitem[\protect\citeauthoryear{{Lin}}{{Lin}}{2019}]{Lin2019}
{Lin} M.-K.,  2019, \mn@doi [\mnras] {10.1093/mnras/stz701}, 485, 5221

\bibitem[\protect\citeauthoryear{{Lin} \& {Youdin}}{{Lin} \&
  {Youdin}}{2015}]{Lin+Youdin2015}
{Lin} M.-K.,  {Youdin} A.~N.,  2015, {\apj}, 811, 17

\bibitem[\protect\citeauthoryear{{Lovelace}, {Li}, {Colgate}  \&
  {Nelson}}{{Lovelace} et~al.}{1999}]{Lovelace+1999}
{Lovelace} R.~V.~E.,  {Li} H.,  {Colgate} S.~A.,   {Nelson} A.~F.,  1999,
  {\apj}, 513, 805

\bibitem[\protect\citeauthoryear{{Lyra}}{{Lyra}}{2014}]{Lyra2014}
{Lyra} W.,  2014, \apj, 789, 77

\bibitem[\protect\citeauthoryear{{Lyra} \& {Umurhan}}{{Lyra} \&
  {Umurhan}}{2019}]{Lyra+Umurhan2019}
{Lyra} W.,  {Umurhan} O.~M.,  2019, \mn@doi [\pasp] {10.1088/1538-3873/aaf5ff},
  131, 072001

\bibitem[\protect\citeauthoryear{{Manger} \& {Klahr}}{{Manger} \&
  {Klahr}}{2018}]{Manger+Klahr2018}
{Manger} N.,  {Klahr} H.,  2018, \mn@doi [\mnras] {10.1093/mnras/sty1909},
  \href {https://ui.adsabs.harvard.edu/\#abs/2018MNRAS.480.2125M} {480, 2125}

\bibitem[\protect\citeauthoryear{{Marcus}, {Pei}, {Jiang}, {Barranco},
  {Hassanzadeh}  \& {Lecoanet}}{{Marcus} et~al.}{2015}]{Marcus+2015}
{Marcus} P.~S.,  {Pei} S.,  {Jiang} C.-H.,  {Barranco} J.~A.,  {Hassanzadeh}
  P.,   {Lecoanet} D.,  2015, \apj, 808, 87

\bibitem[\protect\citeauthoryear{{Marcus}, {Pei}, {Jiang}  \&
  {Barranco}}{{Marcus} et~al.}{2016}]{Marcus+2016}
{Marcus} P.~S.,  {Pei} S.,  {Jiang} C.-H.,   {Barranco} J.~A.,  2016, \apj,
  833, 148

\bibitem[\protect\citeauthoryear{{Meheut}, {Casse}, {Varniere}  \&
  {Tagger}}{{Meheut} et~al.}{2010}]{Meheut+2010}
{Meheut} H.,  {Casse} F.,  {Varniere} P.,   {Tagger} M.,  2010, \aap, 516, A31

\bibitem[\protect\citeauthoryear{{Mignone}}{{Mignone}}{2014}]{Mignone2014}
{Mignone} A.,  2014, Journal of Computational Physics, 270, 784

\bibitem[\protect\citeauthoryear{{Mignone}, {Bodo}, {Massaglia}, {Matsakos},
  {Tesileanu}, {Zanni}  \& {Ferrari}}{{Mignone} et~al.}{2007}]{Mignone+2007}
{Mignone} A.,  {Bodo} G.,  {Massaglia} S.,  {Matsakos} T.,  {Tesileanu} O.,
  {Zanni} C.,   {Ferrari} A.,  2007, \mn@doi [The Astrophysical Journal
  Supplement Series] {10.1086/513316}, 170, 228

\bibitem[\protect\citeauthoryear{Nelson, Gressel  \& Umurhan}{Nelson
  et~al.}{2013}]{Nelson+2013}
Nelson R.~P.,  Gressel O.,   Umurhan O.~M.,  2013, \mn@doi [Monthly Notices of
  the Royal Astronomical Society] {10.1093/mnras/stt1475}, 435, 2610

\bibitem[\protect\citeauthoryear{{Petersen}, {Julien}  \& {Stewart}}{{Petersen}
  et~al.}{2007a}]{Petersen+2007a}
{Petersen} M.~R.,  {Julien} K.,   {Stewart} G.~R.,  2007a, \apj, 658, 1236

\bibitem[\protect\citeauthoryear{{Petersen}, {Stewart}  \& {Julien}}{{Petersen}
  et~al.}{2007b}]{Petersen+2007b}
{Petersen} M.~R.,  {Stewart} G.~R.,   {Julien} K.,  2007b, \apj, 658, 1252

\bibitem[\protect\citeauthoryear{{Raettig}, {Lyra}  \& {Klahr}}{{Raettig}
  et~al.}{2013}]{Raettig+2013}
{Raettig} N.,  {Lyra} W.,   {Klahr} H.,  2013, {\apj}, 765, 115

\bibitem[\protect\citeauthoryear{Rafikov}{Rafikov}{2002}]{Rafikov2002}
Rafikov R.~R.,  2002, The Astrophysical Journal, 569

\bibitem[\protect\citeauthoryear{{Ricci}, {Flock}, {Blanco}  \& {Lyra}}{{Ricci}
  et~al.}{2019}]{Ricci+2019}
{Ricci} L.,  {Flock} M.,  {Blanco} D.,   {Lyra} W.,  2019, arXiv e-prints,
  \href {https://ui.adsabs.harvard.edu/abs/2019arXiv190201897R} {p.
  arXiv:1902.01897}

\bibitem[\protect\citeauthoryear{{Richard}, {Nelson}  \& {Umurhan}}{{Richard}
  et~al.}{2016}]{Richard+2016}
{Richard} S.,  {Nelson} R.~P.,   {Umurhan} O.~M.,  2016, \mn@doi [\mnras]
  {10.1093/mnras/stv2898}, 456, 3571

\bibitem[\protect\citeauthoryear{{R{\"u}diger}, {Arlt}  \&
  {Shalybkov}}{{R{\"u}diger} et~al.}{2002}]{Ruediger+2002}
{R{\"u}diger} G.,  {Arlt} R.,   {Shalybkov} D.,  2002, \aap, 391, 781

\bibitem[\protect\citeauthoryear{{Sch{\"a}fer}, {Johansen}  \&
  {Banerjee}}{{Sch{\"a}fer} et~al.}{2020}]{Schaefer+2020}
{Sch{\"a}fer} U.,  {Johansen} A.,   {Banerjee} R.,  2020, \mn@doi [\aap]
  {10.1051/0004-6361/201937371}, p.~A190

\bibitem[\protect\citeauthoryear{{Shakura} \& {Sunyaev}}{{Shakura} \&
  {Sunyaev}}{1973}]{Shakura+Sunyaev1973}
{Shakura} N.~I.,  {Sunyaev} R.~A.,  1973, Astronomy and Astrophysics, 24, 337

\bibitem[\protect\citeauthoryear{Sharma, Verma  \& Chakraborty}{Sharma
  et~al.}{2018}]{Sharma+2018}
Sharma M.~K.,  Verma M.~K.,   Chakraborty S.,  2018, \mn@doi [Physics of
  Fluids] {10.1063/1.5051444}, 30, 115102

\bibitem[\protect\citeauthoryear{{Stoll} \& {Kley}}{{Stoll} \&
  {Kley}}{2014}]{Stoll+Kley2014}
{Stoll} M.~H.~R.,  {Kley} W.,  2014, \mn@doi [\aap]
  {10.1051/0004-6361/201424114}, 572, A77

\bibitem[\protect\citeauthoryear{{Stoll} \& {Kley}}{{Stoll} \&
  {Kley}}{2016}]{Stoll+Kley2016}
{Stoll} M.~H.~R.,  {Kley} W.,  2016, \mn@doi [\aap]
  {10.1051/0004-6361/201527716}, 594, A57

\bibitem[\protect\citeauthoryear{{Stoll}, {Kley}  \& {Picogna}}{{Stoll}
  et~al.}{2017}]{Stoll+Kley2017}
{Stoll} M.~H.~R.,  {Kley} W.,   {Picogna} G.,  2017, \aap, 599, L6

\bibitem[\protect\citeauthoryear{Toro}{Toro}{2009}]{Toro2009}
Toro E.~F.,  2009, Riemann Solvers and Numerical Methods for Fluid Dynamics.
Springer-Verlag Berlin Heidelberg

\bibitem[\protect\citeauthoryear{{Turner}, {Fromang}, {Gammie}, {Klahr},
  {Lesur}, {Wardle}  \& {Bai}}{{Turner} et~al.}{2014}]{Turner+2014}
{Turner} N.~J.,  {Fromang} S.,  {Gammie} C.,  {Klahr} H.,  {Lesur} G.,
  {Wardle} M.,   {Bai} X.-N.,  2014, Protostars and Planets VI, pp 411--432

\bibitem[\protect\citeauthoryear{{Urpin}}{{Urpin}}{2003}]{Urpin2003}
{Urpin} V.,  2003, \aap, 404, 397

\bibitem[\protect\citeauthoryear{Weidenschilling}{Weidenschilling}{1977}]{Weidenschilling1977}
Weidenschilling S.~J.,  1977, Monthly Notices of the Royal Astronomical
  Society, 180, 57

\makeatother
\end{thebibliography}

\newpage
\onecolumn
\appendix
\section{Azimuthal resolution study }
To determine the minimum grid resolution in the azimuthal direction required to achieve a converged nonlinear state of the VSI and form vortices, we performed a small resolution study. For all simulations we used the parameters of the simulation run p1.5h0.1 described in table \ref{tab:simParam} and varied only $n_\phi$. The simulations performed are listed in table \ref{tab:simNphiParam}.

\begin{table}
\renewcommand\arraystretch{1.5}
\centering
\begin{tabular}{c c c}
\hline
model & grid size ($\mathrm{N}_\mathrm{r}\times\mathrm{N}_\theta \times \mathrm{N}_\varphi$) & $\langle \alpha \rangle /10^{-4}$ \\ \hline
n128  & 256$\times$128$\times$128 & $ 4.9\pm 1.0 $ \\
n256  & 256$\times$128$\times$256 & $5.4\pm 1.3 $ \\
n512  & 256$\times$128$\times$512 & $6.8\pm1.6 $ \\
n1024  & 256$\times$128$\times$1024 & $7.2\pm 1.5$ \\
\hline
\end{tabular}
\caption{List of parameters for the resolution study. The simulation n1024 is identical to p1.5h0.1 from table \ref{tab:simParam}.}
\label{tab:simNphiParam}
\end{table}

\begin{figure*}
    \centering
    \subfloat{\includegraphics[width=0.4\columnwidth]{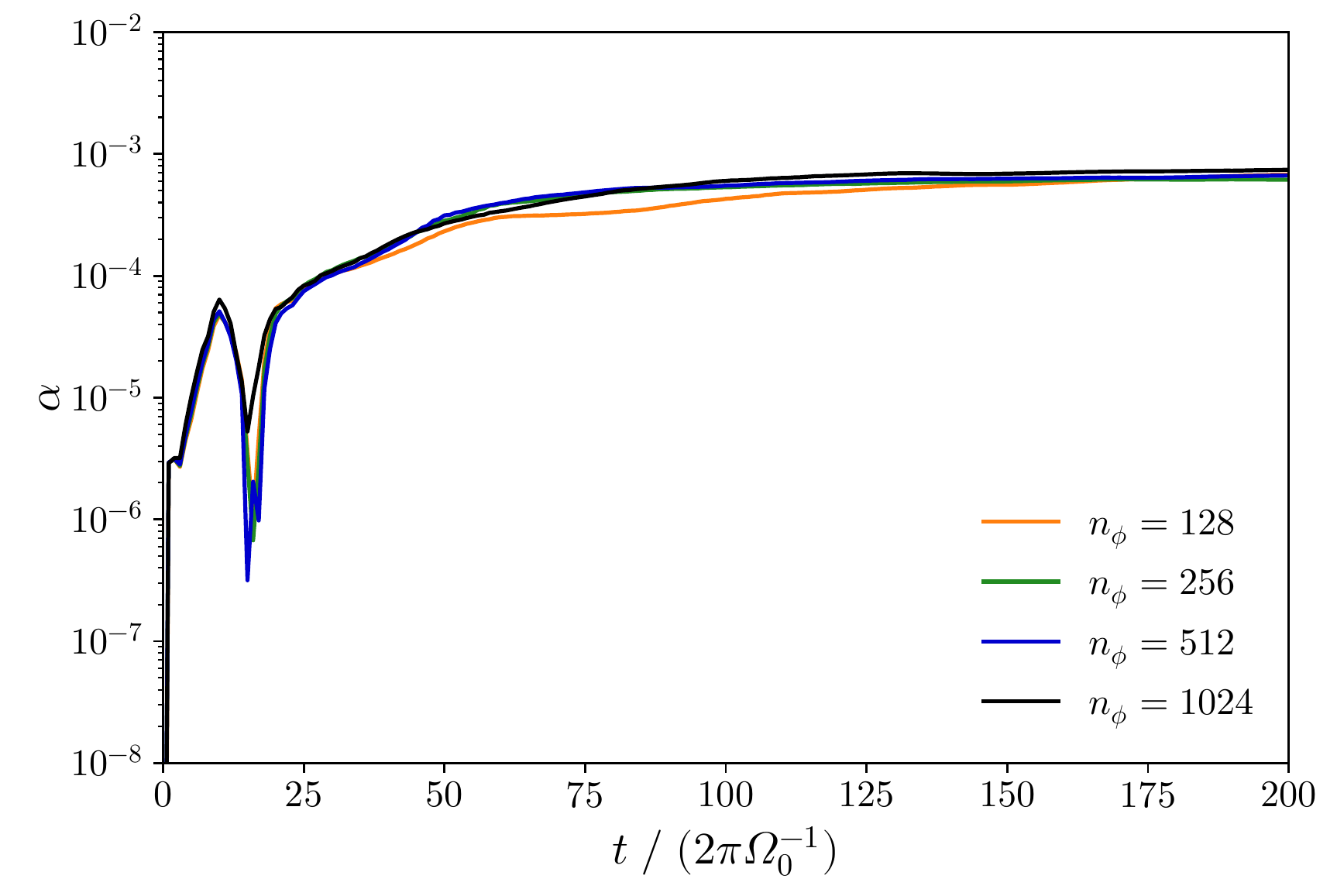}}
    \subfloat{\includegraphics[width=0.4\columnwidth]{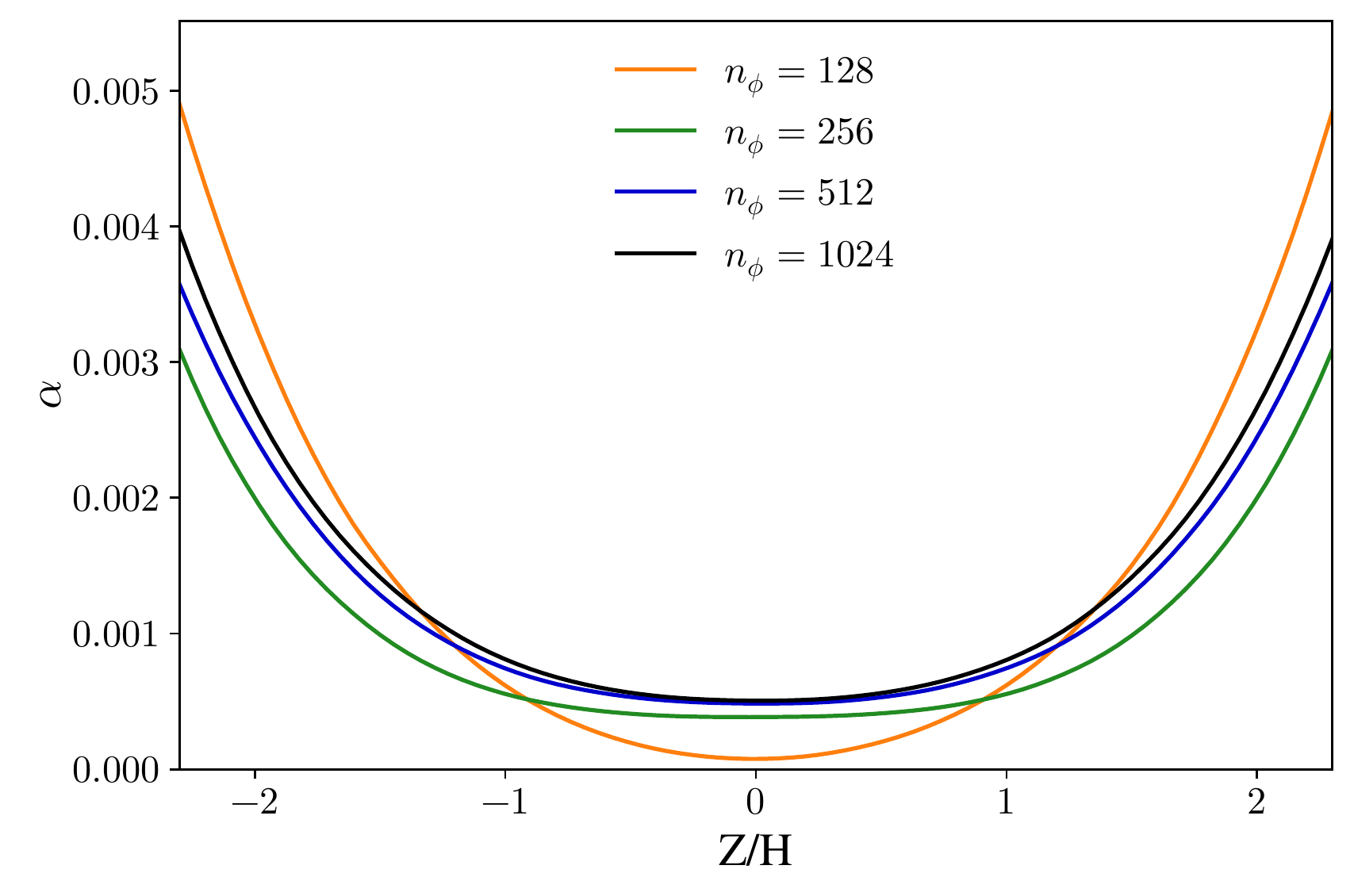}}\\
    \subfloat{\includegraphics[width=0.4\columnwidth]{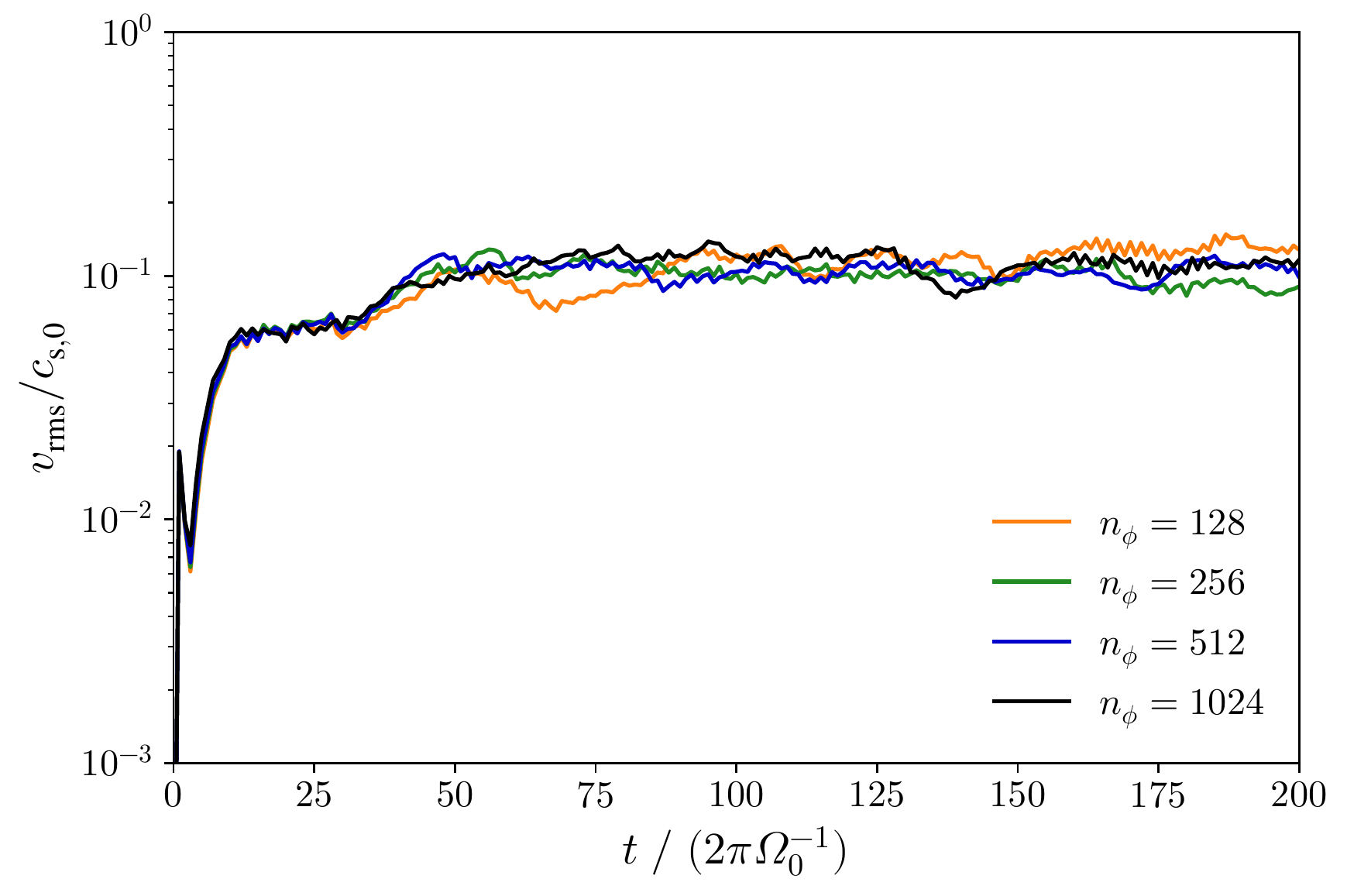}}
    \subfloat{\includegraphics[width=0.4\columnwidth]{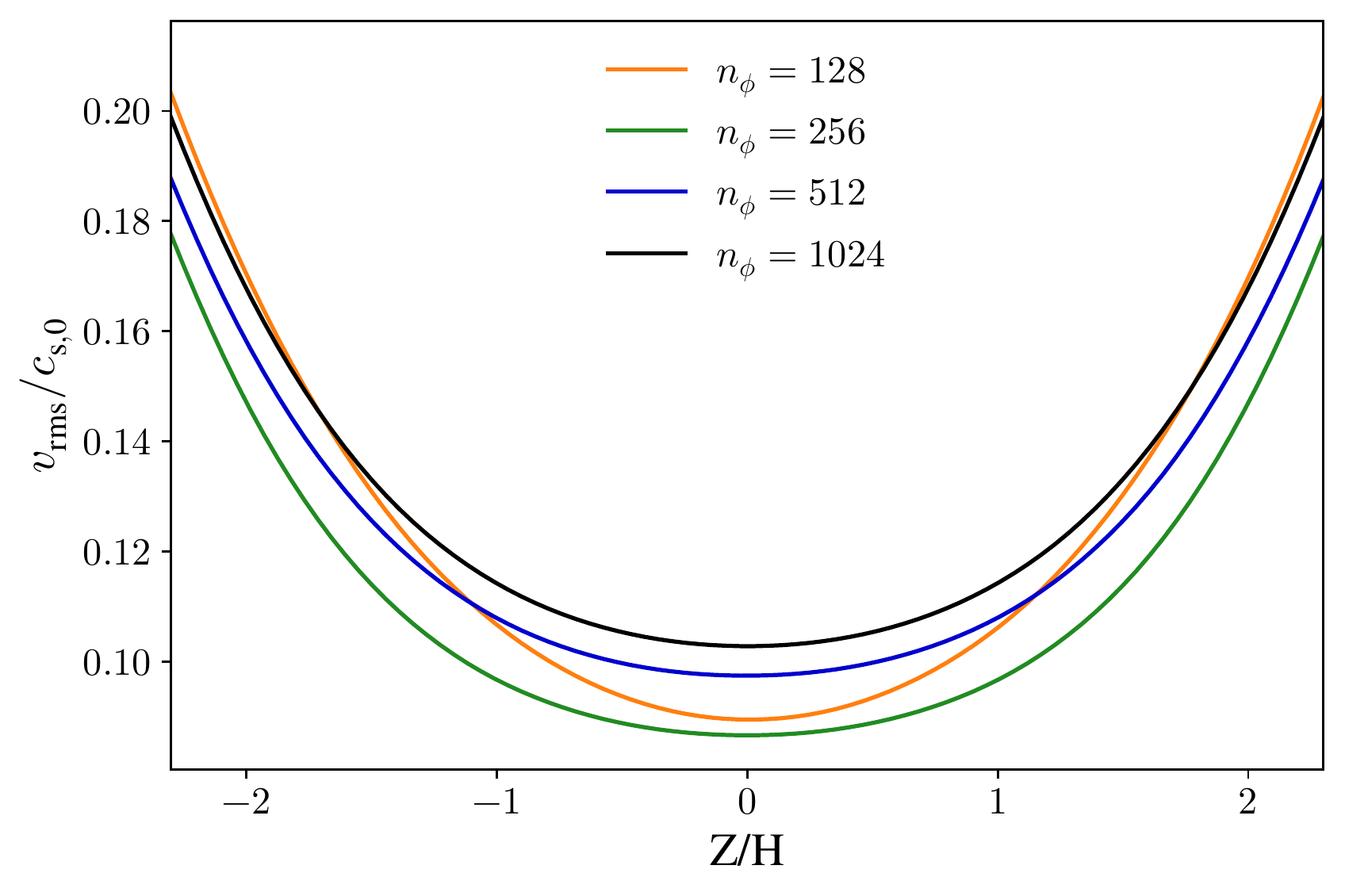}}
    \caption{Turbulence analysis of the simulations performed for the resolution study. The top row shows the $\alpha$ value as a function of time on the left and the  the dependence on height on the right. The bottom row shows the same for the rms-velocity.}
    \label{fig:alphaVrmsNphi}
\end{figure*}

Figure \ref{fig:alphaVrmsNphi} shows the results of the parameter study. We find that both the spatial average of $\alpha$ and $v_\mathrm{rms}$ show converged values which is the expected outcome as the VSI modes grow axisymmetrically. Therefore the formation of vortices does not seem to influence the overall strength of the turbulent angular momentum transport.

\begin{figure}
    \centering
    \includegraphics[width=0.6\columnwidth]{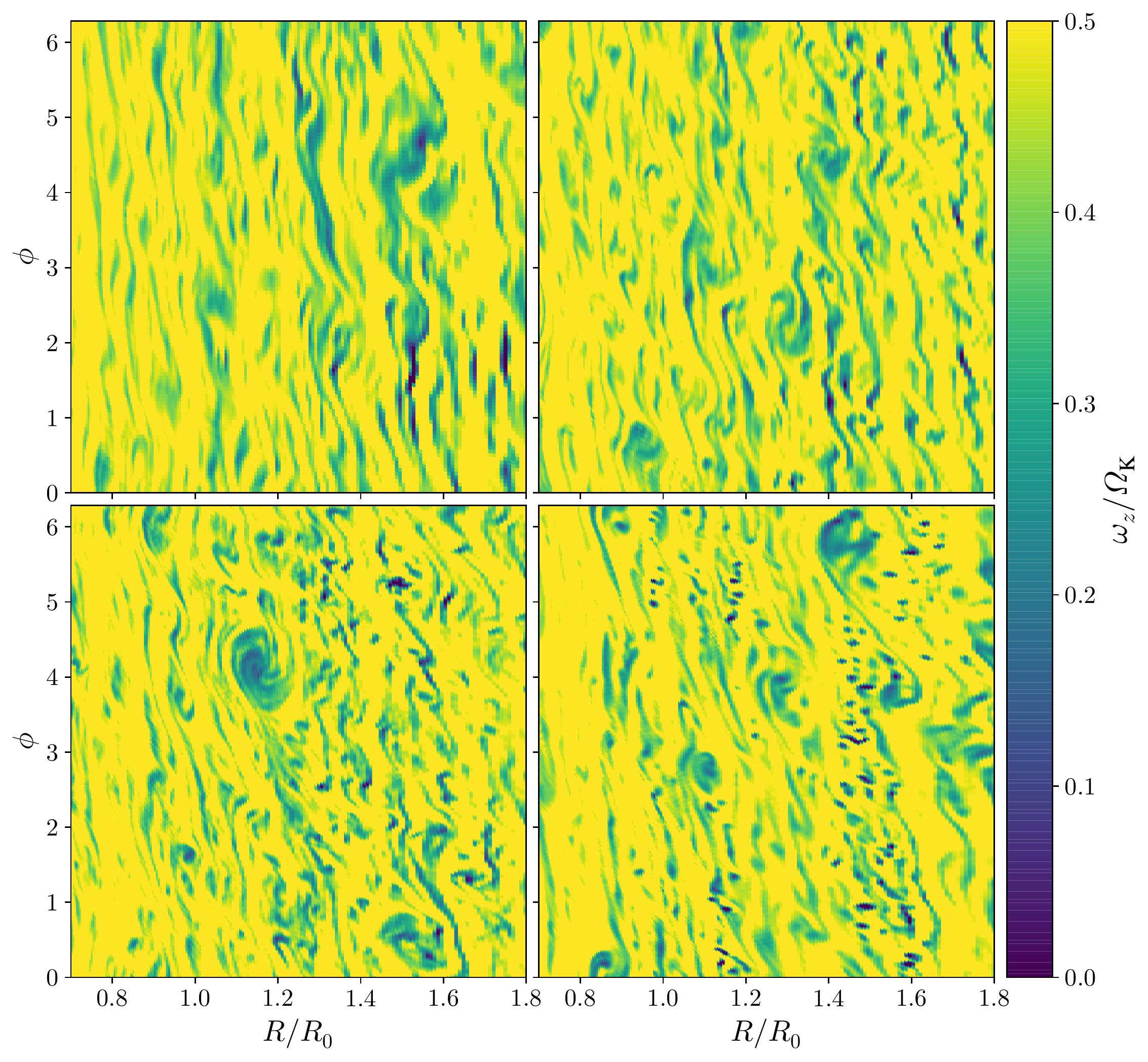}
    \caption{Midplane vorticity after 500 orbits for the simulations considered in the resolution study. Clockwise from top left the phi resolutions are 128, 256, 1024 and 512.}
    \label{fig:vortNphi}
\end{figure}

We also find that the cases n256, n512 and n1024 give similar results when looking at the vertical profiles of $\alpha$ and $v_\mathrm{rms}$. Contrary to this, the simulation n128 shows steeper profile of both values with height than the other simulations. Looking at figure \ref{fig:vortNphi}, we find that n128 is the only case in which no vortex formation is observed. We find a similar behaviour in our analysis presented in \citetalias{Manger+Klahr2018} for the $\alpha$ value for different spatial sizes in the azimuthal direction, where the profile is also steeper for the case where we exclude that vortices form in the disk. Therefore, the formation of vortices could have an influence on where the bulk of the angular momentum transport is facilitated with respect to the midplane of the disk. 

\section{Secondary Rossby Wave Instability}
We plot the Rossby Wave Instability criterion from equation \ref{eqn:RWI} as a function of radius in figure B1 for the simulations with $p=-1.5$ after 200 and 500 orbits. We show the existence of multiple extrema, some of which align well with the radial location of the vortices highlighted in figure \ref{fig:vortZoomH} (dashed red lines). 
\begin{figure*}
    \centering
    \includegraphics[width=\linewidth]{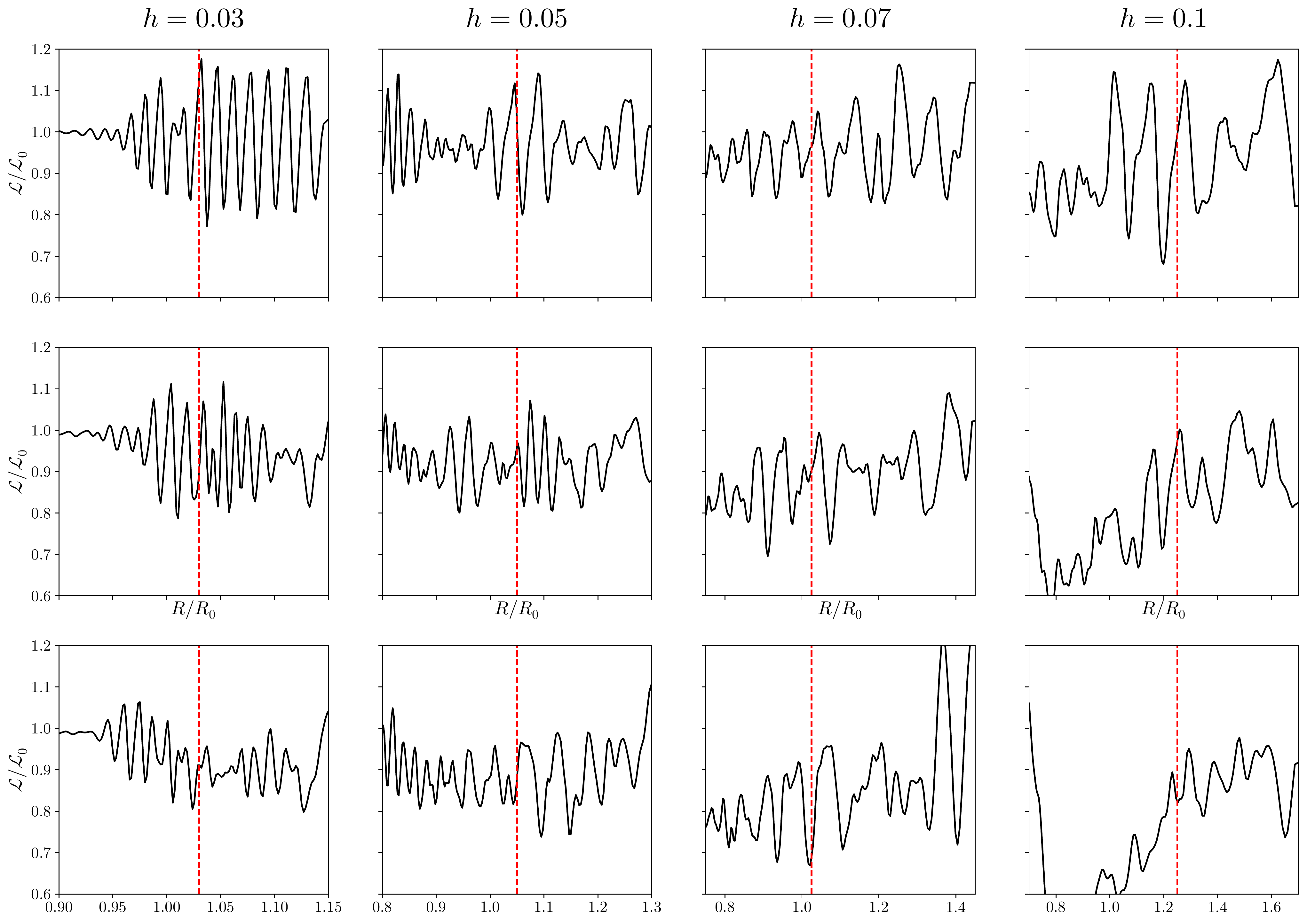}
    \caption{Instability criterion for the RWI \citep{Lovelace+1999} (see eq. \ref{eqn:RWI} in the text) for different $h$. The values are averaged in vertical and azimuth, and show the simulations after 200 (top row), 500 orbits (middle row) and 700 orbits (bottom row). The red lines indicate the positions of the vortices shown in figure \ref{fig:vortZoomH}.}
    \label{fig:RossbyL}
\end{figure*}

\label{lastpage}
\end{document}